\documentclass{ptephy}
\usepackage[latin1]{inputenc}
\usepackage{bm}
\usepackage{amsmath}
\usepackage{amssymb}
\usepackage{graphicx}
\usepackage{esint}

\def\be{\begin{equation}}
\def\ee{\end{equation}}
\def\ba{\begin{eqnarray}}
\def\ea{\end{eqnarray}}

\def\Mpl{M_{\rm pl}}

\renewcommand{\S}{{\text{\tiny $\phi$}}}

\usepackage{color}

\begin{document}

\title{Distinguishing between inflationary models from CMB}

\author{Shinji Tsujikawa}

\address{{Department of Physics, Faculty of Science, Tokyo University of Science, 
1-3, Kagurazaka, Shinjuku-ku, Tokyo 162-8601, Japan, }
\email{shinji@rs.kagu.tus.ac.jp}}

\begin{abstract}

In this paper, inflationary cosmology is reviewed, paying particular attention to 
its observational signatures associated with large-scale density perturbations 
generated from quantum fluctuations.
In the most general scalar-tensor theories with second-order equations 
of motion, we derive the scalar spectral index $n_s$, the tensor-to-scalar 
ratio $r$, and the nonlinear estimator $f_{\rm NL}$ of primordial 
non-Gaussianities to confront models with observations 
of Cosmic Microwave Background (CMB) temperature anisotropies. 
Our analysis includes models such as potential-driven slow-roll inflation, 
k-inflation, Starobinsky inflation, and Higgs inflation with 
non-minimal/derivative/Galileon couplings. 
We constrain a host of inflationary models by 
using the Planck data combined with other measurements 
to find models most favored observationally in the current literature.
We also study anisotropic inflation based on a scalar
coupling with a vector (or, two-form) field and 
discuss its observational signatures appearing in the two-point 
and three-point correlation functions of scalar and tensor perturbations.

\end{abstract}


\maketitle

\section{Introduction}

The inflationary paradigm was first proposed in the early 1980s to 
address the horizon, flatness, and monopole problems that 
plagued Big Bang cosmology \cite{Sta80,oldinf}.
Moreover, inflation provides a causal mechanism for the generation 
of large-scale density perturbations from the quantum 
fluctuation of a scalar field (inflaton).
In its simplest form the resulting power spectra of scalar and 
tensor perturbations are nearly scale-invariant and Gaussian \cite{oldper}.
This prediction showed good agreement with the temperature anisotropies 
of CMB measured by the Cosmic Background Explorer (COBE) \cite{COBE} 
and the Wilkinson Microwave Anisotropy Probe (WMAP) \cite{WMAP1}.
In March 2013 the Planck team \cite{Planck} released more accurate 
CMB data up to the multipoles $\ell \sim 2500$.
With these new data it is now possible to discriminate between a host of 
inflationary models. 

The first model of inflation, proposed by Starobinsky \cite{Sta80}, is 
based on a conformal anomaly in quantum gravity.
The Lagrangian density $f(R)=R+R^2/(6M^2)$, where $R$ 
is a Ricci scalar and $M$ is a mass scale of the order of $10^{13}$ GeV, 
can lead to a sufficient amount of inflation with 
a successful reheating \cite{Starreheating}. 
Moreover, the Starobinsky model is favored from 
the 1-st year Planck observations \cite{Planck}.
The ``old inflation'' \cite{oldinf}, 
which is based on the theory of 
supercooling during the cosmological phase transition, turned 
out to be unviable, because the Universe becomes inhomogeneous 
as a result of the bubble collision after inflation.
The revised version dubbed ``new inflation'' \cite{Linde82,Albrecht}, 
where the second-order transition to true vacuum 
is responsible for cosmic acceleration, 
is plagued by a fine-tuning problem for spending enough time 
in false vacuum. However, these pioneering ideas opened up 
a new paradigm for the construction of workable inflationary models 
based on theories beyond the Standard Model of particle physics 
(see e.g., Refs.~\cite{LRreview,Lindebook,Bau,Mazumdar}).

Most of the inflationary models, including chaotic 
inflation \cite{Linde83}, are based on a slow-rolling scalar field 
with a sufficiently flat potential. 
One can discriminate between a host of inflaton potentials
by comparing theoretical predictions of the scalar spectral 
index $n_s$ and the tensor-to-scalar ratio $r$ 
with the CMB temperature anisotropies 
(see, e.g., \cite{Kolb,LLbook,BTW,Martin}).
The Planck data, combined with the WMAP large-angle polarization (WP) 
measurement, placed the bounds $n_s=0.9603 \pm 0.0073$ (68\,\%\,CL) 
and $r<0.11$ (95\,\%\,CL) for the pivot wavenumber 
$k_0 = 0.002$ \,Mpc$^{-1}$ \cite{Adeinf}. 
Based on the paper \cite{Kuro}, we review the observational bounds 
on potential-driven slow-roll inflation constrained from the joint data analysis of
Planck \cite{Planck}, WP \cite{WMAP9}, Baryon Acoustic 
Oscillations (BAO) \cite{BAO}, and high-$\ell$ \cite{highl}\footnote{Our review
does not reflect constraints derived from the BICEPS2 data for the  
B-mode polarization, released in March 2014.}

Besides slow-roll inflation, there is another class of models, called 
k-inflation models \cite{kinflation}, in which the non-linear field kinetic energy 
plays a crucial role in driving cosmic acceleration. 
Since the scalar propagation speed $c_s$ in k-inflation is generally 
different from the speed of light \cite{Garriga}, this can give rise to large 
non-Gaussianities of primordial perturbations for the equilateral 
shape in the regime $c_s^2 \ll 1$ \cite{Seery,Chen}. 
Using the recent Planck bound on the equilateral non-linear parameter
$f_{\rm NL}^{\rm equil}=-42 \pm 75$ (68\,\%\,CL) \cite{Adenon}, 
it is possible to put tight constraints on most of the k-inflationary models.

There are also other single-field inflationary scenarios constructed
in the framework of extended theories of gravity, such as 
non-minimally coupled models \cite{FM,Higgs}, 
Brans-Dicke theories \cite{Brans}, Galileons \cite{Nico,Vik,KobaGa,Galileons}, 
field derivative couplings to gravity \cite{Amendola,Germani}, and 
running kinetic couplings \cite{Nakayama,Reza}.
All of these models are covered in Horndeski's most general 
scalar-tensor theories with second-order equations of 
motion \cite{Horndeski,DGSZ}. 
For single-field inflation based on the Horndeski theory,
the two-point and three-point correlation functions of 
scalar and tensor perturbations have been computed 
in Refs.~\cite{Koba11,Gao,DT,Kobaten,Gao2} 
(see also Refs.~\cite{nonGa}).
We shall first review these results and then apply them to 
concrete models of inflation.

The WMAP5 data indicated that there is an anomaly associated with the
broken rotational invariance of the CMB power spectrum \cite{aniobser1}. 
This statistical anisotropy is difficult to address
in the context of single-field slow-roll inflation.
The power spectrum of curvature perturbations with broken statistical 
isotropy involves an anisotropy parameter $g_*$. 
This parameter was constrained as $g_*=0.29\pm0.031$ (68\,\%\,CL) from
the WMAP5 data by including multipoles up to $\ell=400$ \cite{aniobser2}. 
With the WMAP9 data, the bound $-0.046<g_*<0.048$ (68\,\%\,CL) was 
derived in Ref.~\cite{Rub}.
Recently, Kim and Komatsu obtained the bound $g_*=0.002\pm0.016$ (68\,\%\,CL)
from the Planck data by taking into account the beam correction 
and the Galactic foreground correction \cite{Kim}.
This result is consistent with the isotropic power spectrum, but 
there is still a possibility that anisotropy of the order $|g_*| \sim 0.01$ remains.

For the models in which the inflaton field $\phi$ has a coupling to 
a vector kinetic term $F_{\mu \nu}F^{\mu \nu}$, an anisotropic hair can 
survive during inflation for a suitable choice of  
coupling $f^2(\phi)$ \cite{Watanabe,Sodareview}. 
In this case, it is possible to explain the anisotropic power spectrum 
compatible with the broken rotational invariance of 
the CMB perturbations \cite{Gum,WataSoda}.
The same property also holds for the two-form field models in which 
the inflaton couples to the kinetic term 
$H_{\mu \nu \lambda}H^{\mu \nu \lambda}$ \cite{Ohashi}, 
but the types of anisotropies are different from each other.
Moreover, these two anisotropic inflationary models can 
give rise to a detectable level of primordial 
non-Gaussianities \cite{Bartolo,Shiraishi,Ohashi,Ohashi2}.
We shall review the general properties of anisotropic inflation and 
discuss their observational signatures.

This review is organized as follows.
In Sec.\,\ref{inspesec} we derive the two-point and three-point 
correlation functions of curvature perturbations and the resulting 
CMB observables in the Horndeski theory.
In Sec.\,\ref{con1sec} we study observational constraints on 
potential-driven slow-roll inflation in the light of the Planck data.
In Sec.\,\ref{con2sec} we distinguish between a host of 
single-field inflationary models that belong to 
the framework of the Horndeski theory. 
In Sec.\,\ref{anisec} we discuss the current status of anisotropic 
inflation paying particular attention to their observational signatures.
Sec.~\ref{consec} is devoted to the conclusion.

\section{Inflationary power spectra and non-Gaussianities 
in the most general scalar-tensor theories}
\label{inspesec}

For generality we start with the action of the most general scalar-tensor theories
with second-order equations of motion \cite{Horndeski,DGSZ,Koba11}
\be
S = \int d^{4}x \sqrt{-g}\
\biggl[\frac{M_{\rm pl}^{2}}{2}\, R+P(\phi,X)
-G_{3}(\phi,X)\,\Box\phi 
+L_{4}+L_{5} \biggr]\,,
\label{action}
\ee
where $g$ is a determinant of the metric tensor 
$g_{\mu\nu}$, $M_{{\rm pl}}$ is the reduced Planck mass, 
$R$ is the Ricci scalar, and 
\ba
\hspace{-1cm}&&L_4=G_4(\phi, X)R+G_{4,X}(\phi, X)
\left[ (\square \phi)^2-\phi^{;\mu \nu}\phi_{;\mu \nu} \right],\\
\hspace{-1cm}&&
L_5 = G_5(\phi,X)G_{\mu \nu}\phi^{;\mu \nu}
-\frac16 G_{5,X} (\phi,X) [ (\square \phi)^3 -3(\square \phi)\,
\phi_{;\mu \nu} \phi^{;\mu \nu}
+2\phi_{;\mu \nu} \phi^{;\mu \lambda} 
{\phi^{;\nu}}_{;\lambda} ]\,.
\ea
Here, a semicolon represents a covariant derivative, 
$P$ and $G_{i}$ ($i=3,4,5$) are functions in terms of $\phi$
and $X\equiv -\partial^{\mu}\phi\partial_{\mu}\phi/2$, and 
$G_{\mu\nu}=R_{\mu\nu}-g_{\mu\nu}R/2$ is the Einstein tensor 
($R_{\mu\nu}$ is the Ricci tensor). 
For the partial derivatives with respect to $\phi$ and $X$,
we use the notation 
$G_{i,\phi} \equiv \partial G_{i}/\partial \phi$ and
$G_{i,X}\equiv \partial G_{i}/\partial X$.

On the flat Friedmann-Lema\^{i}tre-Robertson-Walker (FLRW) background 
described by the line element $ds^2=-dt^2+a^2(t) \delta_{ij}dx^idx^j$, 
the Friedmann equation and the scalar-field equation of motion are given, 
respectively, by \cite{Koba11,Gao,DT}
\ba
&  & 3\Mpl^{2}H^{2}F=P_{,X}\dot{\phi}^2-P
-(G_{{3,\phi}}-12\,{H}^{2}G_{{4,X}}+9\,{H}^{2}G_{{5,\phi}} ){\dot{\phi}^{2}}
-6HG_{{4,\phi}}\dot{\phi}\nonumber \\
&  &
-(6\, G_{{4,\phi X}}-3\, G_{{3,X}}-5\, G_{{5,X}}{H}^{2} )
H{\dot{\phi}^{3}}-3\left(G_{{5,\phi X}}-2\, G_{{4,{\it XX}}}\right)
H^{2}\dot{\phi}^{4}
+{H}^{3}G_{{5,{\it XX}}}{\dot{\phi}^{5}}\,,\label{Friedmann}\\
& &\frac{1}{a^{3}}\frac{d}{dt}\left(a^{3}J\right)=P_{\phi}\,,\label{fieldeq}
\ea
where $H=\dot{a}/a$ is the Hubble parameter (a dot represents 
a derivative with respect to $t$), $F=1+2G_4/M_{\rm pl}^2$, and 
\begin{eqnarray}
J & \equiv & \dot{\phi}P_{,X}+6HXG_{3,X}-2\dot{\phi}G_{3,\phi}
+6H^{2}\dot{\phi}\left(G_{4,X}+2XG_{4,XX}\right)-12HXG_{4,\phi X}\nonumber \\
 &  & +2H^{3}X\left(3G_{5,X}+2XG_{5,XX}\right)-6H^{2}\dot{\phi}
 \left(G_{5,\phi}+XG_{5,\phi X}\right)\,,\\
P_{\phi} & \equiv & P_{,\phi}-2X (G_{3,\phi\phi}+\ddot{\phi}\, G_{3,\phi X} )
+6(2H^{2}+\dot{H})G_{4,\phi}+6H (\dot{X}+2HX )G_{4,\phi X}\nonumber \\
 &  & -6H^{2}XG_{5,\phi\phi}+2H^{3}X\dot{\phi}\, G_{5,\phi X}\,.
\end{eqnarray}

Inflation can be realized in the regime where the slow-roll parameter 
$\epsilon \equiv -\dot{H}/H^2$ is much smaller than 1.
On using Eqs.~(\ref{Friedmann}) and (\ref{fieldeq}), it follows that 
\be
\epsilon=\delta_{PX}+3\delta_{G3X}-2\delta_{G3\phi}+6\,\delta_{G4X}
-\delta_{G4\phi}-6\,\delta_{G5\phi}+3\,\delta_{G5X}+12\,\delta_{G4XX}
+2\,\delta_{G5XX}+O (\epsilon^2)\,,
\label{epsi}
\ee
where the slow-variation parameters on the r.h.s. are 
defined by  $\delta_{PX}=P_{,X}X/(\Mpl^{2}H^{2}F)$, 
$\delta_{G3X}=G_{3,X}\dot{\phi}X/(\Mpl^{2}HF)$, 
$\delta_{G3 \phi}=G_{3,\phi}X/(\Mpl^{2}H^{2}F)$, 
$\delta_{G4X}=G_{4,X}X/(\Mpl^{2}F)$, \\
$\delta_{G4\phi}=G_{4,\phi}\dot{\phi}/(\Mpl^{2}HF)$, 
$\delta_{G5\phi}=G_{5,\phi}X/(\Mpl^{2}F)$, 
$\delta_{G5X}=G_{5,X}H\dot{\phi}X/(\Mpl^{2}F)$, \\
$\delta_{G4XX}=G_{4,XX}X^{2}/(\Mpl^{2}F)$, and 
$\delta_{G5XX}=G_{5,XX}H\dot{\phi}X^{2}/(\Mpl^{2}F)$.

The number of e-foldings is defined as $N(t)=\ln a(t_f)/a(t)$, 
where $a(t)$ and $a(t_f)$ are the scale factors at time $t$ 
during inflation and at the end of inflation respectively.
On using the relation $dN/dt=-H(t)$, it can also be
expressed as
\be
N(t)=-\int_{t_f}^t H(\tilde{t})\,d\tilde{t}\,,
\label{efold}
\ee
where $t_f$ is known by the relation $\epsilon(t_f)=1$.
The number of e-foldings when the perturbations relevant 
to the CMB temperature anisotropies cross the Hubble 
radius is typically in the range $50<N<60$ \cite{Leach,Adeinf}.

For the computations of the two-point and three-point correlation functions 
of scalar and tensor perturbations, we use the following perturbed ADM 
metric \cite{ADM} on the flat FLRW background
\be
ds^2=-[ (1+\alpha)^2-a^{-2}(t) e^{-2\psi} (\partial B)^2] dt^2
+2\partial_{i} B dt dx^i+a^2(t)(e^{2\psi}\delta_{ij}+h_{ij}) dx^i dx^j\,,
\label{permet}
\ee
where $\alpha, B, \psi$ describe scalar metric perturbations, 
and $h_{ij}$ is the tensor perturbation. 
The choice of the ADM metric is particularly convenient 
for the calculation of non-Gaussianities \cite{Maldacena,Koyama}. 
Note that, at linear order in perturbations, the coefficient in front 
of $dt^2$ in Eq.~(\ref{permet}) reduces to $-(1+2\alpha)$.
We introduce the gauge-invariant 
curvature perturbation \cite{Lukash}
\be
\zeta=\psi-\frac{H}{\dot{\phi}} \delta \phi\,,
\label{zeta}
\ee
where $\delta \phi$ is the perturbation in the field $\phi$.
We choose unitary gauge $\delta \phi=0$ to fix the time 
component of a gauge-transformation vector $\xi^{\mu}$. 
The scalar perturbation $E$, which appears 
in the metric (\ref{permet}) in the form $E_{,ij}$, is gauged 
away, so that the spatial component of $\xi^{\mu}$ 
is fixed (see Refs.~\cite{Bardeen,Kodama} for details 
of the cosmological 
perturbation theory).

Expanding the action (\ref{action}) up to second order in perturbations, 
we can derive the equations of motion for linear perturbations. 
Variations of the second-order action with respect to $\alpha$ and $B$
lead to the Hamiltonian and momentum constraints, respectively, 
by which $\alpha$ and $B$ can be related to the curvature 
perturbation $\zeta$.
Then, the resulting second-order action of scalar perturbations 
reads \cite{Koba11,Gao,DT,Kuro}
\be
S_{s}^{(2)}=\int dt\,d^{3}x\, a^{3}Q_s \left[\dot{\zeta}^{2}
-\frac{c_{s}^{2}}{a^{2}}\,(\partial{\zeta})^{2}\right]\,.
\label{secondaction}
\ee
At leading order in slow-variation parameters we have
\ba
Q_s &=& \Mpl^2 F q_s\,,\label{Qs}\\
q_s &\equiv&
\delta_{PX}+2\delta_{PXX}+6\delta_{G3X}+6\delta_{G3XX}
+6\delta_{G4X}+48 \delta_{G4XX}+24\delta_{G4XXX}  \nonumber \\
& &
+6\delta_{G5X}+14\delta_{G5XX}+4\delta_{G5XXX}
-2\delta_{G3\phi}-6\delta_{G5\phi}\,,\label{qs} \\
\epsilon_s &\equiv& \frac{Q_s c_s^2}{ \Mpl^2 F}
=\delta_{PX}+4\delta_{G3X}+6\delta_{G4X}+20\delta_{G4XX}
+4\delta_{G5X}+4\delta_{G5XX}-2\delta_{G3\phi}-6\delta_{G5\phi},
\label{eps2}
\nonumber \\
\ea
where $\delta_{PXX}=X^2 P_{,XX}/(\Mpl^2 H^2 F)$,
$\delta_{G4XXX}=G_{4,XXX}X^3/(\Mpl^2 F)$, and \\
$\delta_{G5XXX}=G_{5,XXX}H\dot{\phi}X^3/(\Mpl^2 F)$.
{}From Eqs.~(\ref{Qs}) and (\ref{eps2}), the scalar propagation 
speed $c_s$ is given by 
\be
c_s^2=\frac{\epsilon_s}{q_s}\,.
\label{cs2d}
\ee
As we will see later, the tensor ghost is absent for $F>0$.
As long as $q_s>0$ and $\epsilon_s>0$, we can avoid 
the ghost and Laplacian instabilities of scalar perturbations.

We write the curvature perturbation in terms of Fourier 
components, as 
\begin{equation}
\zeta (\tau,{\bm{x}})=\frac{1}{(2\pi)^{3}}\int d^{3}{\bm{k}}\,
\hat{\zeta}(\tau,{\bm{k}})e^{i{\bm{k}}\cdot{\bm{x}}}\,,\qquad 
\hat{\zeta}(\tau,{\bm{k}})=\zeta(\tau,{\bm{k}})a({\bm{k}})
+\zeta^{*}(\tau,{-\bm{k}})a^{\dagger}(-{\bm{k}})\,,
\label{RFourier}
\end{equation}
where $\tau=\int a^{-1} dt$ is the conformal time, 
${\bm k}$ is a comoving wavenumber,  and 
$a({\bm{k}})$ and $a^{\dagger}({\bm{k}})$ are the annihilation
and creation operators, respectively, satisfying the commutation relations
\begin{equation}
[a({\bm{k}}_{1}),a^{\dagger}({\bm{k}}_{2})]
=(2\pi)^{3}\delta^{(3)}({\bm{k}}_{1}-{\bm{k}}_{2})\,,
\qquad [a({\bm{k}}_{1}),a({\bm{k}}_{2})]
=[a^{\dagger}({\bm{k}}_{1}),a^{\dagger}({\bm{k}}_{2})]=0\,.
\end{equation}
Since $\tau=-1/(aH)$ in the de Sitter background, 
the asymptotic past and future correspond to $\tau \to -\infty$ 
and $\tau \to -0$, respectively.
Introducing a field $v=z \zeta$ with $z=a\sqrt{2Q_s}$
the kinetic term in the second-order action (\ref{secondaction})
can be rewritten as $\int d\tau d^3 x\,v'^2/2$, 
where a  prime represents a derivative with respect to $\tau$. 
Hence $v$ is a canonical field that should be quantized.
In Fourier space the field $v$ obeys the differential equation
\begin{equation}
v''+\left(c_{s}^{2}k^{2}-\frac{z''}{z}\right)v=0\,.
\label{veq}
\end{equation}
In the de Sitter background with a slow variation of 
the quantity $Q_s$, we have $z''/z \simeq 2/\tau^{2}$. 
In the asymptotic past ($k\tau\to-\infty$) we choose 
the Bunch-Davies vacuum characterized by the mode function 
$v=e^{-ic_{s}k\tau}/\sqrt{2c_{s}k}$.
Then the solution of Eq.~(\ref{veq}) reads
\begin{equation}
\zeta (\tau, k)=\frac{i\,H\, e^{-ic_{s}k\tau}}
{2(c_{s}k)^{3/2}\sqrt{Q_s}}\,(1+ic_{s}k\tau)\,.
\label{usol}
\end{equation}

The two-point correlation function, some time after the 
Hubble radius crossing, is given by the vacuum expectation value
$\langle 0| \hat{\zeta} (\tau, {\bm k}_1) \hat{\zeta} 
(\tau,{\bm k}_2) | 0 \rangle$
at $\tau \approx 0$.
We define the scalar power spectrum 
${\mathcal P}_{\zeta} (k_1)$, as 
\begin{equation}
\langle 0| \hat{\zeta} (0,{\bm k}_1) \hat{\zeta} (0,{\bm k}_2) | 0 
\rangle=(2\pi^2/k_1^3) {\mathcal P}_{\zeta} (k_1)\,
(2\pi)^3 \delta^{(3)} ({\bm k}_1+{\bm k}_2)\,.
\end{equation}
On using the solution (\ref{usol}), the resulting 
power spectrum of $\zeta$ is
\be
{\mathcal P}_{\zeta}=\frac{H^{2}}{8\pi^{2}
M_{\rm pl}^{2}\epsilon_{s}Fc_{s}} \bigg|_{c_sk=aH}\,,
\label{scalarpower}
\ee
which is evaluated at $c_sk=aH$ (because $\zeta$ is nearly 
frozen for $c_sk<aH$). 
The scalar spectral index reads
\be
n_s-1 \equiv \frac{d\ln {\mathcal P}_{\zeta}}{d\ln k}
\bigg|_{c_{s}k=aH}
=-2\epsilon-\eta_{s}-\delta_F-s\,,
\label{nR}
\ee
where 
\be
\eta_s \equiv \frac{\dot{\epsilon}_s}{H \epsilon_s}\,,\qquad
s \equiv \frac{\dot{c}_s}{Hc_s}\,,\qquad
\delta_F \equiv \frac{\dot{F}}{HF}\,.
\ee
The running spectral index is defined by 
$\alpha_s \equiv d n_s/d \ln k |_{c_{s}k=aH}$, 
which is of the order of $\epsilon^2$ from Eq.~(\ref{nR}).

We can decompose the transverse and traceless tensor 
perturbations into two independent polarization modes, as 
$h_{ij}=h_{+}e_{ij}^{+}+h_{\times} e_{ij}^{\times}$, where 
$e_{ij}^{\lambda}$ (where $\lambda=+,\times$) satisfy the relations 
$e_{ij}^{+} ({\bm k}) e_{ij}^{+} (-{\bm k})^*=2$,
$e_{ij}^{\times} ({\bm k}) e_{ij}^{\times} (-{\bm k})^*=2$,
and $e_{ij}^{+} ({\bm k}) e_{ij}^{\times} (-{\bm k})^*=0$.
The second-order action for $h_{ij}$ reads \cite{Koba11,Gao,DT} 
\begin{equation}
S_{t}^{(2)}=\sum_{\lambda=+,\times}\int dt\, d^{3}x\, a^{3} 
Q_{t}\left[ \dot{h}_{\lambda}^{2}
-\frac{c_{t}^2}{a^2} (\partial h_{\lambda})^2 \right]\,,
\label{ST}
\end{equation}
where 
\begin{eqnarray}
Q_{t} &=& \frac{1}{4} M_{{\rm pl}}^{2}F
(1-4\delta_{G4X}-2\delta_{G5X}+2\delta_{G5\phi})\,,
\label{QT} \\ 
c_{t}^{2} &=& 
1+4\delta_{G4X}+2\delta_{G5X}-4\delta_{G5\phi}
+O (\epsilon^2)\,.
\label{cT}
\end{eqnarray}
Following a similar procedure to that for scalar perturbations, 
we obtain the tensor power spectrum
\be
{\mathcal P}_{h}=\frac{H^{2}}
{2\pi^{2}Q_t c_t^3} \biggl|_{c_tk=aH} \simeq 
\frac{2H^2}{\pi^2 M_{\rm pl}^2 F} \biggl|_{k=aH} \,,
\label{Ph}
\ee
where, in the second approximate equality, we have taken
leading-order terms of Eqs.~(\ref{QT}) and (\ref{cT}).

At the epoch when both $\zeta$ and $h_{\lambda}$ become 
nearly constant during inflation, the tensor-to-scalar ratio 
can be evaluated as 
\be
r=\frac{{\mathcal P}_h}{{\mathcal P}_{\zeta}}
\simeq 16 c_s \epsilon_s\,.
\label{ratio}
\ee
Then the tensor spectral index is given by 
\be
n_t \equiv  \frac{d\ln {\mathcal P}_h}{d\ln k}\bigg|_{k=aH}
=-2\epsilon-\delta_F\,.
\label{nt}
\ee
The tensor running $\alpha_t \equiv d n_t/d \ln k |_{k=aH}$ is
of the order of $\epsilon^2$.
On using Eqs.~(\ref{epsi}) and (\ref{eps2}) as well as the 
relation $\delta_F \simeq 2G_{G4\phi}
+O(\epsilon^2)$, we obtain 
the consistency relation 
\be
r=-8c_s \left( n_t-2\delta_{G3X}-16\delta_{G4XX}
-2\delta_{G5X}-4\delta_{G5XX} \right)\,. 
\label{consistency}
\ee

The three-point correlation function of curvature perturbations
associated with scalar non-Gaussianities has been computed 
in Refs.~\cite{Gao,DT}.
The bispectrum $A_{\zeta}$ is defined by 
\be
\langle \zeta ({\bm k}_1) \zeta ({\bm k}_2) 
\zeta ({\bm k}_3) \rangle
=(2\pi)^7 \delta^{(3)} ({\bm k}_1+{\bm k}_2+{\bm k}_3)
({\mathcal P}_{\zeta})^2 \frac{A_{\zeta} (k_1,k_2,k_3)}
{\prod_{i=1}^3 k_i^3}\,.
\ee
The non-linear estimator, $f_{\rm NL}=(10/3)A_{\zeta}/\sum_{i=1}^3 k_i^3$,
is commonly used to characterize the level of 
non-Gaussianities \cite{KSpergel,Bartolo2,Maldacena}.
In Refs.~\cite{Gao,DT} the leading-order bispectrum 
was derived on the de Sitter background.
Reference \cite{DT13} evaluated the three-point correlation 
function by taking into account all possible 
slow-variation corrections to the leading-order 
term (along the lines of Ref.~\cite{Chen}).
Under the slow-variation approximation where each term 
appearing on the r.h.s. of Eq.~(\ref{epsi}) is much smaller than 1, 
the non-linear estimator in the squeezed limit 
($k_3 \to 0$, $k_1 \to k_2$) reads \cite{DT13}
\be
f_{{\rm NL}}^{{\rm local}}=\frac{5}{12} (1-n_s)\,,
\label{flocal}
\ee
which is consistent with the result of Refs.~\cite{Maldacena,Cremi}.
Since $f_{\rm NL}^{\rm local}=O(\epsilon)$, 
the Planck bound $f_{\rm NL}^{\rm local}=2.7 \pm 5.8$ (68\,\%\,CL) \cite{Adenon} 
is satisfied for all the slow-variation single-field models based on the Horndeski 
theory. There are some non-slow roll models in which the non-Gaussianity 
consistency relation (\ref{flocal}) is violated \cite{violation}, but we do not study such 
specific cases.

In the limit of the equilateral triangle ($k_1=k_2=k_3$), 
the leading-order non-linear parameter is given by \cite{DT13}
\ba
f_{{\rm NL}}^{{\rm equil}} &=&  \frac{85}{324}\left(1-\frac{1}{c_{s}^{2}}\right)
-\frac{10}{81}\frac{\lambda}{\Sigma}+\frac{20}{81\epsilon_{s}}
[\delta_{G3X}+\delta_{G3XX} +4(3\delta_{G4XX}+2\delta_{G4XXX})
+\delta_{G5X} \nonumber \\
& &+5\delta_{G5XX}+2\delta_{G5XXX}] 
+\frac{65}{162c_{s}^{2}\epsilon_{s}}(\delta_{G3X}+6\delta_{G4XX}
+\delta_{G5X}+\delta_{G5XX})\,,
\label{fnleq}
\ea
where 
\begin{eqnarray}
\lambda & = & 
\frac{F^{2}}{3}[3X^{2}P_{,XX}+2X^{3}P_{,XXX}
+3H\dot{\phi}(XG_{3,X}+5X^{2}G_{3,XX}+2X^{3}G_{3,XXX}) \nonumber \\
& &-2(2X^{2}G_{3,\phi X}+X^{3}G_{3,\phi XX})+6H^{2}(9X^{2}G_{4,XX}
+16X^{3}G_{4,XXX}+4X^{4}G_{4,XXXX})\nonumber \\
& & -3H\dot{\phi}(3XG_{4\phi,X}+12X^{2}G_{4,\phi XX}+4X^{3}G_{4,\phi XXX})
+H^{3}\dot{\phi}(3XG_{5,X}+27X^{2}G_{5,XX} \nonumber \\
& &+24X^{3}G_{5,XXX}+4X^{4}G_{5,XXXX}) 
-6H^{2}(6X^{2}G_{5,\phi X}+9X^{3}G_{5,\phi XX}+2X^{4}G_{5,\phi XXX})],\nonumber \\
\Sigma &=& 
\frac{Q_s}{4M_{\rm pl}^4}
[2M_{\rm pl}^{2}HF-2X\dot{\phi}G_{3,X}-16H(XG_{4,X}+X^{2}G_{4,XX})
+2\dot{\phi}(G_{4,\phi}+2XG_{4,\phi X})\nonumber \\
 &  & {}-2H^{2}\dot{\phi}(5XG_{5,X}+2X^{2}G_{5,XX})+4HX(3G_{5,\phi}+2XG_{5,\phi X})]^2\,.
\end{eqnarray}
For the models in which $c_s^2$ is much smaller than 1, the nonlinear estimator 
$|f_{\rm NL}^{\rm equil}|$ can be much larger than 1.
The Planck team derived the bound 
$f_{\rm NL}^{\rm equil}=-42 \pm 75$  (68\,\%\,CL) \cite{Adenon}
by using three optimal bispectrum estimators.
The primordial non-Gaussianities provide additional constraints 
on the models to those derived from $n_s$ and $r$.

\section{Planck constraints on potential-driven 
slow-roll inflation}
\label{con1sec}

Let us first study observational constraints on standard 
slow-roll inflation characterized by the functions
\be
P(\phi, X)=X-V(\phi),\quad
G_3=0\,,\quad G_4=0\,,\quad G_5=0\,,
\label{standinf}
\ee
where $V(\phi)$ is the inflaton potential.
Under the slow-roll approximations $\dot{\phi}^2/2 \ll V$ and 
$|\ddot{\phi}| \ll |3H \dot{\phi}|$, Eqs.~(\ref{Friedmann}) 
and (\ref{fieldeq}) reduce to $3\Mpl^2 H^2 \simeq V$ and 
$3H \dot{\phi} \simeq -V_{,\phi}$ respectively.
Then the number of e-foldings (\ref{efold}) can be 
expressed as
\be
N \simeq \frac{1}{M_{\rm pl}^2} \int_{\phi_f}^{\phi}
\frac{V}{V_{,\tilde{\phi}}} d \tilde{\phi}\,,
\label{efoldstan}
\ee
where $\phi_f$ is the field value at the end of inflation 
known by the condition $\epsilon(\phi_f)=1$.

The slow-roll parameter $\epsilon$ is equivalent to 
$\epsilon_s=\dot{\phi}^2/(2M_{\rm pl}^2 H^2)$.
Under the slow-roll approximation it follows that 
$\epsilon_s \simeq \epsilon_V$ and 
$\eta_s \simeq 4\epsilon_V-2\eta_V$, where
\be
\epsilon_V \equiv \frac{\Mpl^2}{2} 
\left( \frac{V_{,\phi}}{V} \right)^2\,,
\qquad
\eta_V \equiv \frac{\Mpl^2 V_{,\phi \phi}}{V}\,.
\label{epV}
\ee
Using the fact that $c_s^2=1$ and $F=1$, 
the observables (\ref{nR}), (\ref{ratio}), and 
(\ref{nt}) reduce to
\be
n_s=1-6\epsilon_V+2\eta_V\,,\qquad 
r=-8n_t\,,\qquad
n_t=-2\epsilon_V\,.
\label{nsr}
\ee
For a given inflaton potential these observables can be
expressed in terms of $\phi$. 
The field value corresponding to $N=50 \sim 60$ 
is known by Eq.~(\ref{efoldstan}).

For observational constraints on inflationary models 
based on the Planck data, we expand the scalar and tensor 
power spectra around a pivot wavenumber $k_0$, as 
\ba
& &\ln {\mathcal P}_{\zeta} (k)= \ln {\mathcal P}_{\zeta} (k_0)
+\left[ n_{s}(k_0)-1 \right]x+
\alpha_{s}(k_0)x^2/2+O(x^3)\,,
\label{Pexp1} \\
& &\ln {\mathcal P}_h (k)= \ln {\mathcal P}_h (k_0)
+n_t(k_0)x+\alpha_{t}(k_0)x^2/2
+O(x^3)\,,
\label{Pexp2}
\ea
where $x=\ln (k/k_0)$. 
Since the likelihood results are insensitive to the choice 
of $k_0$, we fix $k_0=0.05~{\rm Mpc}^{-1}$ as in Ref.~\cite{Kuro}.
Since the runnings $|\alpha_s (k_0)|$ and $|\alpha_t (k_0)|$ 
are of the order of $\epsilon^2$ under the slow-roll approximation, 
we also set these values to 0. 
Using the consistency relation $r(k_0)=-8n_t(k_0)$, the three 
inflationary observables ${\mathcal P}_{\zeta} (k_0)$, 
$n_s(k_0)$, and $r(k_0)$ are varied in the likelihood analysis.
We also assume the flat $\Lambda$CDM model with 
$N_{\rm eff}=3.046$ relativistic degrees of freedom \cite{Mangano}
and employ the standard Big Bang nucleosynthesis 
consistency relation \cite{Ichi}.
In addition to the Planck data \cite{Adeinf}, we also use
the data of WP \cite{WMAP9}, BAO \cite{BAO}, and 
high-$\ell$ \cite{highl}.

\begin{figure}
\begin{center}
\includegraphics[height=3.8in,width=3.7in]{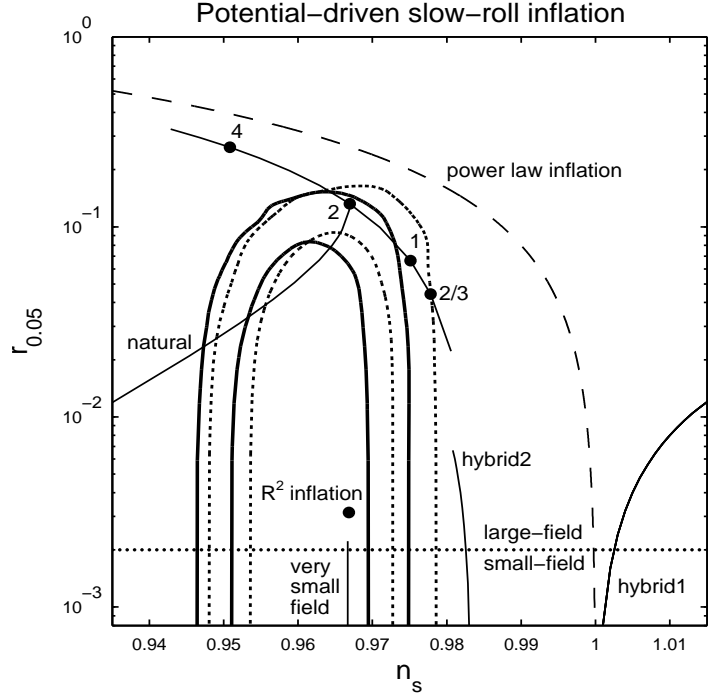}
\end{center}
\caption{\label{fig1}
Observational constraints on potential-driven slow-roll inflation 
in the $(n_s,r)$ plane with $N=60$ and $k_0=0.05$~Mpc$^{-1}$.
The thick solid and dotted curves correspond to
the 68\,\%\,CL (inside) and 95\,\%\,CL (outside) boundaries 
derived by the joint data analysis of Planck+WP+BAO+high-$\ell$ 
and that of Planck+WP+BAO, respectively. 
We show the theoretical predictions for the 
models: (i) chaotic inflation with the potential 
$V(\phi)=\lambda_n \phi^n/n$ for general $n$ (thin solid curve) and 
for $n=4,2,1,2/3$ (denoted by black circles), (ii) natural inflation with 
the potential $V(\phi)=\Lambda^4 [1+\cos (\phi/f)]$ for general $f$, 
(iii) hybrid inflation with the potentials 
$V(\phi)=\Lambda^4+m^2 \phi^2/2$ (``hybrid1'') and 
$V(\phi)=\Lambda^4 [1+c \ln (\phi/\mu)]$ (``hybrid2''), 
(iv) very small-field inflation with the potential 
$V(\phi)=\Lambda^4 (1-e^{-\phi/M})$ in the regime 
$M<M_{\rm pl}$, and 
(v) power-law inflation with the exponential potential 
$V(\phi)=V_0 e^{-\gamma \phi/M_{\rm pl}}$.
The dotted line ($r=2 \times 10^{-3}$) corresponds to the boundary
between ``large-field'' and ``small-field'' models. 
We also show the theoretical prediction of the Starobinsky 
model $f(R)=R+R^2/(6M^2)$ (shown as ``$R^2$ inflation'' 
in the figure).}
\end{figure}

In Fig.~\ref{fig1} we plot the 68\,\%\, and 95\,\%\, CL boundaries
in the $(n_s, r)$ plane constrained by the joint data analyses of 
Planck, WP, BAO, high-$\ell$ data (thick solid curves) and 
Planck, WP, BAO data (thick dotted curves). 
With the high-$\ell$ data, the scalar spectral index shifts toward smaller 
values and  the tensor-to-scalar ratio gets slightly smaller. 
In what follows we place observational constraints on a number of 
representative inflaton potentials. 
For observational bounds on other potentials, 
we refer the reader to Ref.~\cite{Martin}. 
The Planck constraints on braneworld inflation \cite{Maartens} and 
non-commutative inflation \cite{Ho} have been studied in Ref.~\cite{Calcagni} 
for several inflaton potentials discussed below, but we do not have 
any significant observational evidence that they are 
particularly favored over standard slow-roll inflation.

\subsection{Chaotic inflation}

Chaotic inflation is characterized by the potential \cite{Linde82}
\be
V(\phi)=\lambda_n \phi^n/n\,,
\label{powerlaw}
\ee
where $n$ and $\lambda_n$ are positive constants. 
In this case the slow-roll parameters (\ref{epV}) reduce to 
$\epsilon_V=n^2 M_{\rm pl}^2/(2\phi^2)$ and 
$\eta_V=n(n-1)M_{\rm pl}^2/\phi^2$.
{}From Eq.~(\ref{efoldstan}) we obtain the relation
$\phi^2(N) \simeq 2n \left( N+n/4 \right)M_{\rm pl}^2$, 
where we used $\phi_f=nM_{\rm pl}/\sqrt{2}$. 
The scalar spectral index and the tensor-to-scalar ratio read
\be
n_{s}=1-\frac{2(n+2)}{4N+n}\,,\qquad
r=\frac{8n}{n+2} (1-n_s)\,.
\label{nsrstan}
\ee
As we see in Fig.~\ref{fig1}, the quartic potential ($n=4$) is 
far outside the 95\,\%\,CL contour. 
For the quadratic potential ($n = 2$) we have $n_s = 0.967$ and 
$r =0.132$ for $N=60$, which is marginally inside the 95\,\%\,CL boundary 
constrained by the Planck+WP+BAO+high-$\ell$ data.
For $N=50$ the quadratic potential is outside the 95\,\%\,CL region.

The potentials with the powers $n=1$ and $n=2/3$ appear in 
the axion monodromy scenario \cite{Mca,West}.
For $N=60$, the linear potential is within the 95\,\%\,CL region 
constrained by the Planck+WP+BAO data, 
but it is outside the 95\,\%\,CL boundary by adding the high-$\ell$ data. 
For $N=50$ the linear potential enters the joint 95\,\%\,CL region
constrained by the Planck+WP+BAO+high-$\ell$ data due to the 
decrease of $n_s$.
For $N=60$ the potential with $n=2/3$ is outside the 
joint 95\,\%\,CL boundary derived by the 
Planck+WP+BAO+high-$\ell$ data, 
but for $N=50$ the model marginally lies within the 
95\,\%\,CL contour. 

The exponential potential $V(\phi)=V_0 e^{-\gamma \phi/M_{\rm pl}}$ 
corresponds to the limit $n \to \infty$ in Eq.~(\ref{nsrstan}), 
which is characterized by the line $r=8(1-n_s)$.
This model, which is shown as a dashed curve in Fig.~\ref{fig1},  
is excluded at more than 95\,\%\,CL.

\subsection{Natural inflation}

We proceed to natural inflation given by the potential \cite{natural}
\be
V(\phi)=\Lambda^4 \left[ 1+\cos(\phi/f) \right]\,,
\label{naturalpo}
\ee
where $\Lambda$ and $f$ are constants. 
{}From Eq.~(\ref{efoldstan}) the number of e-foldings 
can be estimated as $N \simeq \delta_f^{-1} 
\ln \{(2\epsilon_V+\delta_f)/[(2+\delta_f)\epsilon_V]\}$, 
where $\delta_f=M_{\rm pl}^2/f^2$.
This is inverted to give 
\be
\epsilon_V \simeq \frac{\delta_f}{e^{N\delta_f}(2+\delta_f)-2}\,.
\label{epVna}
\ee
The slow-roll parameter $\eta_V$ is related to $\epsilon_V$ via
\be
\eta_V=\epsilon_V-\delta_f/2\,.
\ee
For given values of $N$ and $f$, we can evaluate
$n_s=1-4\epsilon_V-\delta_f$ and $r=16\epsilon_V$ by using 
Eq.~(\ref{epVna}). 
In the limit that $f \to \infty$, inflation occurs
in the regime where $\phi$ is close to the potential minimum
($\phi=\pi f$). Since $\epsilon_V \to 1/(2N+1)$ and $\delta_f \to 0$
in this limit, we obtain $n_s=1-4/(2N+1)$ and $r=16/(2N+1)$, which 
correspond to the values of chaotic inflation with $n=2$. 

In Fig.~\ref{fig1}, we plot the theoretical values of $n_s$ 
and $r$ for $N=60$ as a function of $f$.
For decreasing $f$, both $n_s$ and $r$ get smaller.
{}From the joint analysis of the Planck+WP+BAO+high-$\ell$ data,  
the symmetry-breaking scale $f$ is constrained as
\be
5.1M_{\rm pl}<f<7.9M_{\rm pl} \qquad  (68\,\%\,{\rm CL})\,,
\label{naturalcon}
\ee
whereas $f>4.6 M_{\rm pl}$ at 95\,\%\,CL.

\subsection{Hybrid inflation}

Hybrid inflation involves two scalar fields: the inflaton $\phi$ and 
another symmetry-breaking field $\chi$. 
During inflation the field $\chi$ is close to 0, in which 
regime the potential is approximately given by 
\be
V(\phi) \simeq \Lambda^4 +U(\phi)\,,
\label{hybridpo}
\ee
where $\Lambda$ is a constant, and $U(\phi)$
depends on $\phi$. Inflation ends due to 
a waterfall transition driven by the growth of $\chi$.
Linde's original hybrid model \cite{hybrid} 
corresponds to $U(\phi)=m^2 \phi^2/2$. Provided that 
the ratio $r_U \equiv U(\phi)/\Lambda^4$ is much smaller 
than 1, it follows that 
\be
n_s \simeq 1+\frac{2m^2 M_{\rm pl}^2}{\Lambda^4}\,,\qquad
r \simeq 8(n_s-1) r_{U}\,.
\ee
Hence the scalar power spectrum is blue-tiled ($n_s>1$).
Under the condition $r_U<0.1$, the tensor-to-scalar ratio is 
bounded as $r<0.8(n_s-1)$ (which is shown as 
a solid curve in Fig.~\ref{fig1} in the regime $n_s>1$).
The hybrid model with $U(\phi)=m^2 \phi^2/2$ is far outside 
the 95\,\%\,CL region.

There is also a supersymmetric GUT model characterized by the 
potential $V(\phi)=\Lambda^4+c\,\Lambda^4 \ln (\phi/\mu)$ 
with $c \ll 1$ \cite{Dvali}. In the regime where $\phi$ is much larger than 
the field value $\phi_c$ at the bifurcation point, 
the observables are given by $n_s \simeq 1-(2+3c)/(2N)$
and $r \simeq 4c/N$, where we have used 
$N \simeq \phi^2/(2M_{\rm pl}^2 c)$.
Since the second derivative $V_{,\phi \phi}$ is negative, 
the spectrum is red-tilted. In Fig.~\ref{fig1} the theoretical curves 
are plotted for $0<c<0.1$ and $N=60$.
The model is outside the 95\,\%\,CL region constrained by 
the Planck+WP+BAO+high-$\ell$ data
due to the large scalar spectral index.

\subsection{Very small-field inflation}

The tensor-to-scalar ratio is related to the variation of the field
during inflation.
In fact, we obtain the relation $(d\phi/dN)^2 \simeq M_{\rm pl}^2r/8$ 
from Eqs.~(\ref{efoldstan})-(\ref{nsr}).
Provided that $r$ is nearly constant during inflation, the 
field variation $\Delta \phi$ is approximately given by \cite{Lyth,Lyth2}
\be
\frac{\Delta \phi}{M_{\rm pl}} \simeq 
\left( \frac{r}{2 \times 10^{-3}} \right)^{1/2} 
\left( \frac{N}{60} \right)\,.
\label{Lythbound}
\ee
The models with $\Delta \phi<M_{\rm pl}$ are dubbed small-field inflation, 
in which case $r$ is smaller than $2 \times 10^{-3}$ for $N=60$.
In Ref.~\cite{Kuro} the criterion $r=10^{-2}$ was used to separate large-field and 
small-field models. Here we employ a more precise criterion 
according to Eq.~(\ref{Lythbound}).

Small-field inflation can be realized by the potential
\be
V(\phi)=\Lambda^4 [1-\mu(\phi)]\,,
\ee
where $\Lambda$ is a constant and
$\mu(\phi)$ is a function of $\phi$. 
In D-brane inflation \cite{Tye} and K\"{a}hler-moduli 
inflation \cite{Conlon} we have $\mu(\phi)=e^{-\phi/M}$ and 
$\mu(\phi)=c_1 \phi^{4/3}e^{-c_2\phi^{4/3}}$ 
($c_1>0$, $c_2>0$), respectively.
See Refs.~\cite{KKLT} for other similar models.

For the function $f(\phi)=e^{-\phi/M}$ the number of e-foldings 
is given by $N \simeq (M/M_{\rm pl})^2 e^{\phi/M}$, in which 
case $n_s$ and $r$ are
\be
n_s \simeq 1-\frac{2}{N}\,,\qquad  r \simeq \frac{8}{N^2}
\left( \frac{M}{M_{\rm pl}} \right)^2\,.
\ee
For $M<M_{\rm pl}$ and $N=60$, it follows that 
$n_s \simeq 0.967$ and $r<2.2 \times 10^{-3}$. 
The model is inside the 68\,\%\,CL boundary 
constrained by the Planck+WP+BAO+high-$\ell$ data.

In K\"{a}hler-moduli inflation the inflationary observables are
in the ranges $0.960<n_s<0.967$ and $r<10^{-10}$
for $50<N<60$ \cite{Conlon}, so the model belongs 
to a class of very small-field inflation.
This model is consistent with the observational data as well.

\section{Discrimination between general single-field 
models from observations}
\label{con2sec}

In this section we study observational constraints on single-field 
inflationary models that belong to the class of the Horndeski
theory given by the action (\ref{action}).

\subsection{k-inflation}

Kinetically driven inflation--called k-inflation \cite{kinflation}--corresponds
to the action 
\be
S=\int d^4x \sqrt{-g} \left[ \frac{M_{\rm pl}^2}{2}R+P(\phi,X) \right]\,,
\ee
i.e., $G_3=G_4=G_5=0$ in (\ref{action}). 
Since $\epsilon=\epsilon_s=\delta_{PX}=XP_{,X}/(\Mpl^2 H^2)$,
inflation occurs around either $X \approx 0$ or $P_{,X} \approx 0$.
The former corresponds to slow-roll inflation discussed in 
Sec.~\ref{con1sec}, whereas the latter is k-inflation driven by 
the presence of non-linear terms in $X$.

{}From Eq.~(\ref{cs2d}) the field propagation 
speed squared is given by  \cite{Garriga}
\be
c_s^2=\frac{P_{,X}}{P_{,X}+2XP_{,XX}}\,.
\ee
The observables (\ref{nR}) and (\ref{ratio}) reduce to
\be
n_s-1=-2\epsilon-\eta_s-s\,,\qquad
r=16c_s \epsilon\,.
\ee
Since $\lambda/\Sigma=(1-c_s^2)/2+2X^2P_{,XXX}c_s^2/(3P_{,X})$ \cite{DT}, 
the equilateral leading-order non-linear parameter (\ref{fnleq}) reads
\be
f_{\rm NL}^{\rm equil}=\frac{5}{324} \left( 1-\frac{1}{c_s^2} \right)
(17+4c_s^2)-\frac{20}{243} \frac{X^2P_{,XXX}}{P_{,X}+2XP_{,XX}}\,.
\ee
If $c_s^2 \ll 1$, then $|f_{\rm NL}^{\rm equil}|$ can be much larger than 1.

To be concrete, we discuss the dilatonic ghost condensate 
model \cite{ghost2} characterized by the Lagrangian 
\be
P(\phi, X)=-X+e^{\alpha \phi/M_{\rm pl}}X^2/M^4\,,
\label{dila}
\ee
where $\alpha$ and $M$ are constants. 
When $\alpha=0$, this recovers the ghost 
condensate model \cite{ghost}.
Since $\epsilon=3(2Xe^{\alpha \phi/M_{\rm pl}}-M^4)/
(3Xe^{\alpha \phi/M_{\rm pl}}-M^4)$, 
$c_s^2=(2Xe^{\alpha \phi/M_{\rm pl}}-M^4)/(6Xe^{\alpha \phi/M_{\rm pl}}-M^4)$, 
and $P_{,XXX}=0$, the inflationary observables can be expressed as
\be
n_s-1=n_t=-\frac{24c_s^2}{1+3c_s^2}\,,\qquad
r=\frac{192c_s^3}{1+3c_s^2}\,,\qquad
f_{\rm NL}^{\rm equil}=\frac{5}{324} \left( 1-\frac{1}{c_s^2} \right)
(17+4c_s^2)\,.
\label{nsrdbi}
\ee
{}From the joint data analysis based on $n_s$ and $r$, 
the sound speed is constrained as
$0.034<c_s<0.046$ (95\,\%\,CL) from the Planck+WP+BAO+high-$\ell$ 
data \cite{Kuro}. Using an equilateral template of primordial 
non-Gaussianities, the Planck team derived the bound $c_s>0.079$ (95 \,\%\,CL).
Hence the dilatonic ghost condensate model is disfavored by 
adding constraints from non-Gaussianities to those 
derived from $n_s$ and $r$.

Let us consider the Dirac-Born-Infeld (DBI) model \cite{DBIinf,DBIsky} 
characterized by the Lagrangian
\be
P(\phi, X)=-f(\phi)^{-1} \sqrt{1-2f(\phi)X}+f(\phi)^{-1}-V(\phi)\,,
\label{DBIac}
\ee
where $f(\phi)$ is a warp factor and $V(\phi)$ is a field potential.
Since $\lambda/\Sigma=(1-c_s^2)/(2c_s^2)$ in this case, the equilateral 
non-linear estimator (\ref{fnleq}) reads
\be
f_{\rm NL}^{\rm equil}=\frac{35}{108}
\left(1 -\frac{1}{c_s^2} \right)\,.
\ee
Using this relation, the Planck team derived 
the bound $c_s>0.07$ (95 \,\%\,CL) \cite{Adenon}. 
The ultraviolet DBI models \cite{DBIinf} correspond to the functions 
$f(\phi) \propto \phi^{-4}$ and $V(\phi)=m^2 \phi^2/2$.
Taking into account the bounds from $n_s$ and $r$, this model is
incompatible with the data for theoretically consistent 
model parameters \cite{Baumann,Adeinf}.
In the infrared DBI model characterized by 
$f(\phi)\propto \phi^{-4}$ and $V(\phi)=V_0-\beta H^2 \phi^2/2$ \cite{IR} ($0.1<\beta<10^9$) 
the joint constraints from $n_s$, $r$, and $f_{\rm NL}^{\rm equil}$ restrict
the allowed parameter space in a narrow range: $\beta<0.7$ (95\,\%\,CL).
In the power-law DBI model with the functions 
$f(\phi)^{-1} \propto e^{-\gamma \phi/M_{\rm pl}}$ and 
$V(\phi) \propto e^{-\gamma \phi/M_{\rm pl}}$ \cite{assistedk}, 
the scalar propagation speed is contained as
$0.07<c_s<0.43$ (95\,\%\,CL) from the bounds 
of $n_s$, $r$, and $f_{\rm NL}^{\rm equil}$ \cite{Kuro}.

\subsection{Starobinsky inflation}

Let us consider the so-called $f(R)$ gravity described by the action 
\be
S=\int d^4 x \sqrt{-g}\,\frac{M_{\rm pl}^2}{2}f(R)\,,
\ee
where $f(R)$ is an arbitrary function of $R$. 
This can be written as 
\be
S=\int d^4 x \sqrt{-g}\left[ \frac12 M_{\rm pl} \phi R
-V(\phi) \right]\,,\quad {\rm where} \quad
\frac{\phi}{M_{\rm pl}}=\frac{\partial f}{\partial R}\,,\quad
V(\phi)=\frac{M_{\rm pl}^2}{2} \left( R \frac{\partial f}{\partial R}
-f \right)\,.
\label{fR2}
\ee
Provided the function $f(R)$ includes non-linear terms of $R$,  
the scalar degree of freedom $\phi$ propagates. 
The field has a potential $V(\phi)$ of gravitational origin.
We write (\ref{fR2}) in a more general form
\be
S=\int d^4 x \sqrt{-g}\left[ \frac{M_{\rm pl}^2}{2} F(\phi) R
+\omega(\phi)X-V(\phi) \right]\,,
\label{fRge}
\ee
where $f(R)$ gravity corresponds to $F(\phi)=\phi/M_{\rm pl}$
and $\omega(\phi)=0$ \cite{Ohanlon}. 
Under the conformal transformation $\hat{g}_{\mu \nu}=F(\phi)g_{\mu \nu}$ 
we obtain the following action in the Einstein frame \cite{Maeda}:
\begin{equation}
\hat{S}=\int d^{4}x\sqrt{-\hat{g}}\left[\frac{1}{2}\Mpl^{2}\hat{R}
-\frac{1}{2}\hat{g}^{\mu\nu}\partial_{\mu}\chi\partial_{\nu}\chi
-U(\chi)\right]\,,
\label{Eaction}
\end{equation}
where a hat represents quantities in the Einstein frame, and
\begin{equation}
U=\frac{V}{F^{2}}\,,\qquad \chi=\int {\mathcal B}(\phi)\, d\phi\,,\qquad 
{\mathcal B}(\phi)=\sqrt{\frac{3}{2}\left(\frac{\Mpl F_{,\phi}}{F}\right)^{2}
+\frac{\omega}{F}}\,.
\end{equation}

Let us consider the Starobinsky model \cite{Sta80}
\begin{equation}
f(R)=R+R^2/(6M^2)\,.
\end{equation}
In this case the field potential in the Einstein frame is 
given by 
\begin{equation}
U(\chi)=\frac34 M^2 M_{\rm pl}^2 \left( 1-
e^{-\sqrt{2/3}\,\chi/M_{\rm pl}} \right)^2\,,\quad
{\rm where} \quad
\chi=\sqrt{\frac32}M_{\rm pl} \ln \left( 1
+\frac{R}{3M^2} \right)\,.
\end{equation}
In the limit that $\chi \to \infty$ the potential approaches a constant
$U(\chi) \to 3M^2M_{\rm pl}^2/4$, so that inflation occurs in the regime 
$\chi \gg M_{\rm pl}$.
The slow-roll parameters $\hat{\epsilon}_V$ and $\hat{\eta}_V$ in 
this regime can be estimated as 
$\hat{\epsilon}_V \simeq 3/(4N^2)$ and 
$\hat{\eta}_V \simeq -1/N$, where $N \simeq (3/4)e^{\sqrt{2/3}\chi/M_{\rm pl}}$ 
is the number of e-foldings. In Refs.~\cite{equiva} it was shown that the inflationary 
power spectra of scalar and tensor perturbations 
in the original (Jordan) frame are equivalent to those in 
the Einstein frame. {}From Eq.~(\ref{nsr}) the inflationary 
observables are \cite{fRper}
\begin{equation}
n_s=1-\frac{2}{N}\,,\qquad
r=\frac{12}{N^2}\,.
\end{equation}
When $N=60$ we have $n_s=0.967$ and $r=0.0033$.
As we see in Fig.~\ref{fig1}, the Starobinsky model is well inside
the 68\,\%\,CL contour constrained by the 
Planck+WP+BAO+high-$\ell$ data.
See Refs.~\cite{Stainf} for theoretical attempts to construct the Starobinsky 
model in the framework of supergravity and quantum gravity.

\subsection{Higgs inflation}

From the amplitude of the CMB anisotropies the typical mass scale of inflation 
is around $H \sim 10^{14}$ GeV \cite{Adeinf}. This is much higher than the electroweak 
scale ($\sim 10^{2}$ GeV), so the Higgs field cannot be responsible for inflation 
in its simplest form. However, this situation is subject to change in the presence 
of non-minimal couplings or other interactions.
In what follows we briefly review a number of approaches to 
accommodate the Higgs field for inflation and discriminate those 
models from observations.

\subsubsection{Non-minimal couplings}

The models with non-minimal couplings between the inflaton 
and the Ricci scalar are described by the action \cite{FM,Higgs} 
\begin{equation}
S=\int d^4 x \sqrt{-g} \left[ \frac{M_{\rm pl}^2}{2}R
-\frac12 \xi \phi^2 R+X-V(\phi) \right]\,,
\label{nonmo}
\end{equation}
where the conformal coupling corresponds to $\xi=1/6$. 
The model (\ref{nonmo}) corresponds to $F(\phi)=1-\xi \phi^2/M_{\rm pl}^2$
and $\omega(\phi)=1$ in (\ref{fRge}). Then the action in the Einstein frame is 
given by (\ref{Eaction}) with the potential $U=V/F^2$. 
The Higgs potential $V(\phi)=(\lambda_4/4)(\phi^2-v^2)^2$ ($v \sim 10^2$ GeV) 
can be approximated as $V(\phi) \simeq \lambda_4 \phi^4/4$ 
during inflation ($\phi^2 \gg v^2$).
Then the potential in the Einstein frame reads 
\begin{equation}
U \simeq \frac{\lambda_4 \phi^4}{4(1-\xi \phi^2/M_{\rm pl}^2)^2}\,.
\end{equation}
For negative $\xi$, the potential is asymptotically flat 
in the regime $-\xi \phi^2/M_{\rm pl}^2 \gg 1$.

Let us consider the case of a large negative non-minimal 
coupling ($|\xi| \gg 1$). The scalar power spectrum 
(\ref{scalarpower}) can be estimated as 
${\mathcal P}_{\zeta} \simeq \lambda_4 N^2/(72\pi^2 \xi^2)$ with 
$N \simeq -(3/4)\xi \phi^2/M_{\rm pl}^2$.
Using the Planck normalization 
${\mathcal P}_{\zeta}=2.2 \times 10^{-9}$ \cite{Adeinf,Kuro}
with $N=60$, we obtain $\lambda_4/\xi^2 \simeq -4.3 \times 10^{-10}$.
For the non-minimal coupling $\xi \approx -10^{4}$, the 
self coupling $\lambda_4$ can be of the order of 0.1.
Since the slow-roll parameters in the Einstein frame 
are given by $\epsilon_V \simeq (4/3)(M_{\rm pl}^2/(\xi \phi^2))^2$ 
and $\eta_V \simeq 4M_{\rm pl}^2/(3\xi \phi^2)$ \cite{Gumjud}, we obtain 
\begin{equation}
n_s \simeq 1-\frac{2}{N}\,,\qquad 
r \simeq \frac{12}{N^2} \qquad (|\xi| \gg 1).
\label{nsrhi}
\end{equation}
Provided that quantum corrections to the tree-level action are
suppressed, the theoretical values (\ref{nsrhi})
are the same as those in the Starobinsky model, 
so the model is within the 68\,\%\,CL observational contour. 

A detailed analysis shows that the non-minimal coupling is constrained 
as $\xi<-4.5 \times 10^{-3}$ (68\,\%\,CL) from the joint data analysis 
of Planck+WP+BAO+high-$\ell$ \cite{Kuro}.

\subsubsection{Field-derivative couplings to the Einstein tensor}

Let us proceed to the field-derivative coupling model 
described by the action \cite{Germani}
\be
S = \int d^{4}x \sqrt{-g}\
\left[\frac{M_{\rm pl}^{2}}{2} R+X-V(\phi)
+\frac{1}{2M^2} G^{\mu \nu} \partial_{\mu} \phi
\partial_{\nu} \phi \right]\,,
\label{fieldde}
\ee
where $M$ is a constant having a dimension of mass.
In the regime where the Hubble parameter $H$ is larger 
than $M$, the field evolves more slowly relative to the case 
of standard inflation due to a gravitationally enhanced 
friction\footnote{This property is similar to what happens for warm 
inflation \cite{Berera}, in which dissipative processes lead to 
an effective friction for the inflaton.}.

For a slow-rolling field satisfying the 
condition $\varepsilon=\dot{\phi}^2/(M^2M_{\rm pl}^2) \ll 1$, 
the strong coupling scale $\Lambda_c$ of the derivative 
coupling theory is as close as the Planck scale 
$M_{\rm pl}$ \cite{Germani}.
This comes from the fact that an asymptotic local shift 
symmetry (related to the Galilean symmetry mentioned 
later in Sec.~\ref{Galisec}) is only softly broken 
for $\varepsilon \ll 1$, so that the potential can be protected 
against quantum corrections during inflation even 
in the regime $M<H<M_{\rm pl}$ \cite{Germani2}. 
Note that the sign of the last term of Eq.~(\ref{fieldde}) 
has been chosen to avoid the appearance of ghosts.

\begin{figure}
\begin{center}
\includegraphics[height=3.4in,width=3.4in]{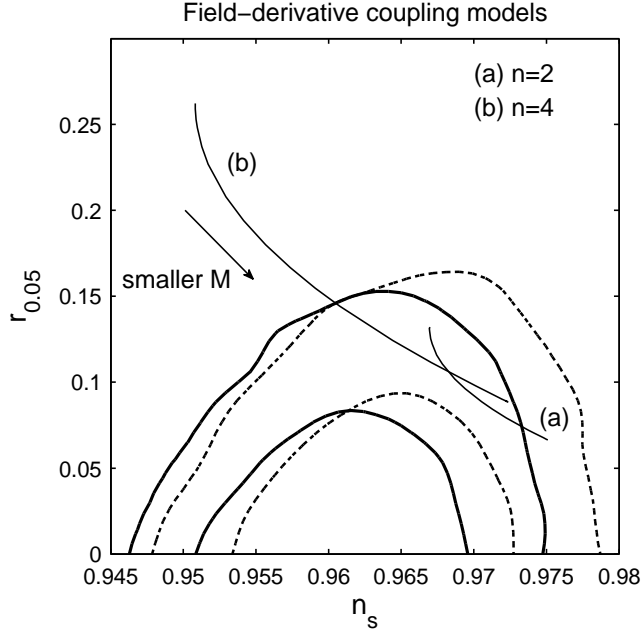}
\end{center}
\caption{\label{fig2}
Observational constraints on field-derivative coupling models (\ref{fieldde}) 
for the monomial potential $V(\phi)=\lambda_n \phi^n/n$. 
The thick solid and dotted curves show the 68\,\%\,CL (inside) 
and 95\,\%\,CL (outside) boundaries derived by the joint data 
analyses of Planck+WP+BAO+high-$\ell$ and Planck+WP+BAO, respectively. 
The thin solid curves correspond to the theoretical predictions
of the models $n=2$ and $n=4$, respectively (for $N=60$).
For decreasing $M$ the scalar spectral index gets larger, 
whereas the tensor-to-scalar ratio becomes smaller.}
\end{figure}

To be concrete, let us consider the monomial potential 
given in (\ref{powerlaw}).
In this case, Eqs.~(\ref{nR}) and (\ref{ratio}) read \cite{Wata,tsujideri}
\begin{equation}
n_s=1-\frac{n^2[n(n+2)+2(n+1)\alpha y^n]}
{y^2 (n+\alpha y^n)^2}\,,\qquad
r=\frac{8n^3}{y^2 (n+\alpha y^n)}\,,
\end{equation}
where $\alpha=\lambda_n M_{\rm pl}^{n-2}/M^2$ and 
$y=\phi/M_{\rm pl}$. 
The number of e-foldings is given by 
$N=y^2[1+2\alpha y^n/(n(n+2))]/(2n)-
y_f^2[1+2\alpha y_f^n/(n(n+2))]/(2n)$, where 
$y_f$ is known by solving $2y_f^2(1+\alpha y_f^n/n)=n^2$.
In the limit $\alpha \to \infty$, it follows that 
\be
n_s=1-\frac{4(n+1)}{2(n+2)N+n}\,,\qquad
r=\frac{16n}{2(n+2)N+n}\,.
\label{deriasy}
\ee
If $N=60$, then $n_s=0.975$, $r=0.066$ for $n=2$
and $n_s=0.972$, $r=0.088$ for $n=4$.

In Fig.~\ref{fig2} we plot theoretical curves in the $(n_s, r)$ plane for 
$n=2$ and $n=4$ in the parameter range $10^{-8} \le \alpha \le 10^8$. 
Although $r$ gets smaller for decreasing $M$ due to the enhanced 
gravitational friction, 
both potentials are outside the 68\,\%\,CL region. 
In the limit $\alpha \to \infty$, the potential $V(\phi)=\lambda_4 \phi^4/4$ is 
marginally inside the 95\,\%\,CL contour. {}From the joint data 
analysis of Planck+WP+BAO+high-$\ell$, the same potential 
is inside the 95\,\%\,CL boundary for $\alpha>9.0 \times 10^{-5}$ \cite{Kuro}.

In the regime $H \gg M$ the scalar power spectrum is approximately 
given by ${\mathcal P}_{\zeta} \simeq V^4/(12\pi^2 M_{\rm pl}^8M^2 V_{,\phi}^2)$.
Using the Planck normalization ${\mathcal P}_{\zeta} \simeq 2.2 \times 10^{-9}$ 
at $N=60$, the self coupling is constrained as
$\lambda_4 \simeq 6 \times 10^{-32} (M_{\rm pl}/M)^4$. 
For $M \simeq 3 \times 10^{-8}M_{\rm pl}$ it is possible to realize 
$\lambda_4 \simeq 0.1$.

\subsubsection{Galileon self-interactions}
\label{Galisec}

The field equations of motion following from the Lagrangian 
$X \square \phi$ are invariant under the Galilean shift
$\partial_{\mu} \phi \to \partial_{\mu} \phi+b_{\mu}$ 
in the limit of Minkowski space-time \cite{Nico}. 
The general covariant Galileons \cite{Vik} having the 
same property as the term $X \square \phi$ 
can be accommodated by the action (\ref{action}) 
with the choice
\be
P=X-V(\phi)\,,\qquad 
G_3=\frac{c_3}{M^3}X\,,\qquad
G_4=-\frac{c_4}{M^6}X^2\,,\qquad
G_5=\frac{3c_5}{M^9}X^2\,,
\label{Galilag}
\ee
where only the linear potential $V(\phi) \propto \phi$ is allowed 
to respect the Galilean symmetry in the limit of Minkowski 
space-time. In the following we do not restrict the form of the 
field potential to the linear one since the Galilean symmetry 
is broken in the curved space-time anyway. 
In the presence of the terms $G_3,G_4,G_5$ given in 
Eq.~(\ref{Galilag}) the evolution of the inflaton 
along the potential also slows down \cite{Kamada}. 
For simplicity, let us consider the case where the terms 
$G_4$ and $G_5$ are absent. 
{}From Eqs.~(\ref{qs}) and (\ref{cs2d}) we have 
\be
q_s=\delta_{PX}+6 \delta_{G3X}\,,\qquad
c_s^2=\frac{\delta_{PX}+4 \delta_{G3X}}
{\delta_{PX}+6 \delta_{G3X}}\,,
\ee
where $ \delta_{G3X}=c_3 \dot{\phi}X/(M^3 M_{\rm pl}^2H)$. 
When $|\delta_{G3X}| \gg \delta_{PX}$ the Galileon 
self-interaction dominates over the standard kinetic term.
In this regime, the avoidance of ghosts requires the condition 
$\delta_{G3X}>0$, i.e., $c_3 \dot{\phi}>0$. 
The propagation speed squared is approximately 
given by $c_s^2 \simeq 2/3$, 
so that the Laplacian instability can be avoided during 
inflation.
Since $r=16(\delta_{PX}+4 \delta_{G3X})^{3/2}
/(\delta_{PX}+6 \delta_{G3X})^{1/2}$ 
and $n_t=-2(\delta_{PX}+3\delta_{G3X})$, the consistency relation 
in the regime $|\delta_{G3X}| \gg \delta_{PX}$ is given by 
$r \simeq -8.7n_t$.

Let us consider the monomial potential (\ref{powerlaw}). 
In the regime where $M$ is much smaller than $H$, 
the observables (\ref{nR}) and (\ref{ratio}) 
reduce to \cite{Kamada,OhashiGa}
\be
n_s=1-\frac{3(n+1)}{(n+3)N+n} \,, \qquad 
r=\frac{64 \sqrt{6}}{9}\frac{n}{(n+3)N+n}\,,
\label{nrgal}
\ee
which give $n_s=0.965$ and $r=0.164$ for 
$n=4$ and $N=60$.
For intermediate values of $M$ 
the tensor-to-scalar ratio of the potential $V(\phi)=\lambda_4 \phi^4/4$ 
is in the range $0.164<r<0.262$ for $N=60$, in which case the model is 
outside the 95  \,\%\,CL boundary constrained by 
the Planck+WP+BAO+high-$\ell$ data \cite{Kuro}. 
For the Galileon couplings $G_4=-c_4X^2/M^6$ or 
$G_5=3c_5X^2/M^9$ the self-coupling potential enters 
the 95\,\%\,CL contour in the presence of Galileon terms, 
but it is still outside the 68\,\%\,CL contour \cite{Kuro}. 
The quadratic potential $V(\phi)=m^2 \phi^2/2$ is also 
outside the 68\,\%\,CL boundary.

If the Galileon term dominates over the standard kinetic term even after inflation, 
this gives rise to instabilities associated with the appearance 
of negative $c_s^2$ during reheating \cite{OhashiGa}. 
This provides a lower bound on the mass scale $M$ of
$M \gtrsim 10^{-4}M_{\rm pl} \approx 10^{14}$ GeV for 
the monomial potential  (\ref{powerlaw}). 
This lower bound is similar to the typical energy scale 
of inflation. Hence the Hubble parameter $H$ is not actually 
much larger than $M$ during inflation. 
Even for $H \sim M$, however, $n_s$ and $r$ are close
 to the values (\ref{nrgal}) \cite{OhashiGa}.

\section{Anisotropic inflation}
\label{anisec}

The WMAP data showed that there may be some 
anomalies related to the broken rotational invariance of 
the CMB perturbations \cite{aniobser1}.
The power spectrum of curvature perturbations 
with broken statistical isotropy can be parametrized as 
\be
{\mathcal P}_{\zeta} ({\bm k})=
{\mathcal P}_{\zeta}^{(0)} (k)\,\left(1+g_*\cos^2 
\theta_{{\bm k},{\bm V}} \right)\,,
\label{anispe}
\ee
where ${\mathcal P}_{\zeta}^{(0)} (k)$ is the isotropic power spectrum, 
$g_*$ quantifies the deviation from the isotropy, ${\bm V}$ is a privileged 
direction close to the ecliptic poles, and $\theta_{{\bm k},{\bm V}}$ is 
the angle between the wavenumbers ${\bm k}$ and ${\bm V}$. 
{}From the WMAP5 data, Groeneboom {\it et al.} \cite{aniobser2} derived 
the bound $g_*=0.29 \pm 0.031$ with the exclusion of $g_*=0$ 
at $9\sigma$. {}From the Planck data the bound 
$g_*=0.002 \pm 0.016$ (68\%\,CL) was recently derived by 
Kim and Komatsu \cite{Kim} after eliminating the asymmetry of 
the beam and the Galactic foreground emission. 
Although the isotropic spectrum is consistent with the Planck data, 
the anisotropy of the order $|g_*|=0.01$ has not yet been excluded.

In order to explain the origin of statistical anisotropies, 
we need to go beyond the simplest single-field inflationary scenario.
If the inflaton couples to a vector field with a kinetic term 
$F_{\mu \nu}F^{\mu \nu}$, the anisotropic hair can survive 
during inflation for a suitable choice of coupling $f^2(\phi)$ \cite{Watanabe}.
In the following we review the mechanism of anisotropic inflation 
and evaluate the anisotropy parameter $g_*$ as well as 
the non-linear parameter $f_{\rm NL}$ 
in such a scenario.

\subsection{Anisotropic hair}

We start with the following action 
\be
S=\int d^4x\sqrt{-g}\left[ \frac{M_{\rm pl}^2}{2}R
+X-V(\phi)-\frac{1}{4} f^2(\phi) F_{\mu\nu}F^{\mu\nu}  
\right]\,,
\label{eq:action}
\ee
where the field strength of the vector field is characterized by
$F_{\mu\nu} = \partial_\mu A_\nu - \partial_\nu A_\mu$.
Choosing the gauge $A_0=0$, the $x$-axis can be taken 
for the direction of the vector, i.e., $A_{\mu}=(0, v(t),0,0)$, 
where $v(t)$ is a function with respect to $t$.
There is rotational symmetry in the $(y,z)$ plane, 
so that the line element can be expressed as
\begin{equation}
ds^2 = -{\mathcal N}^2(t) dt^2 + e^{2\alpha(t)} \left[ e^{-4\sigma (t)}dx^2
+e^{2\sigma (t)}(dy^2+dz^2) \right] \ ,
\label{anisotropic-metric}
\end{equation}
where ${\mathcal N}(t)$ is the lapse function, and
$e^\alpha \equiv a$ and $\sigma$ are the isotropic 
scale factor and the spatial shear, respectively.
Then the action (\ref{eq:action}) can be written as
\be
S=\int d^4 x \frac{e^{3\alpha}}{{\mathcal N}} \left[ 3M_{\rm pl}^2
(\dot{\sigma}^2-\dot{\alpha}^2)+\frac12 \dot{\phi}^2-{\mathcal N}^2V(\phi)
+\frac12 f^2(\phi) e^{-2\alpha+4\sigma} \dot{v}^2 \right]\,.
\label{actioncon}
\ee

The equation of motion for $v$ following from the action (\ref{actioncon}) 
can be integrated to give 
\be
\dot{v}=p_A\,f^{-2}(\phi) e^{-\alpha-4\sigma}\,,
\ee
where $p_A$ is a constant. 
Varying the action (\ref{actioncon}) with respect to 
${\mathcal N}$, $\alpha$, $\sigma$, $\phi$, and setting 
${\mathcal N}=1$, we obtain
\begin{eqnarray}
& &H^2=\dot{\sigma}^2+\frac{1}{3M_{\rm pl}^2}
\left[ \frac12 \dot{\phi}^2+V(\phi)+\rho_A\right]\,,
\label{beani1} \\
& &\dot{H}=-3\dot{\sigma}^2-\frac{1}{M_{\rm pl}^2}
\left( \frac12 \dot{\phi}^2+\frac23 \rho_A \right)\,,
\label{beani2} \\
& & \ddot{\sigma}=-3H \dot{\sigma}+\frac{2\rho_A}
{3M_{\rm pl}^2}\,,
\label{beani3} \\
& &\ddot{\phi}+3H\dot{\phi} 
+V_{,\phi}(\phi)
-p_A^2f^{-3}(\phi) f_{,\phi} (\phi)e^{-4\alpha-4\sigma}=0\,,
\label{beani4}
\end{eqnarray}
where $H=\dot{\alpha}$, and the energy density 
of the vector field is given by 
\be
\rho_A \equiv \frac12 p_A^2 
f^{-2}(\phi)e^{-4\alpha -4\sigma}\,.
\label{rhoA}
\ee
The inflaton energy density $\rho_{\phi} \equiv \dot{\phi}^2/2+V(\phi)$ 
needs to be much larger than $\rho_A$ to sustain inflation.
Moreover, the Hubble parameter $H$ should be much larger than 
the shear term $\Sigma \equiv \dot{\sigma}$, so that 
Eq.~(\ref{beani1}) is approximately given by 
$H^2 \simeq \rho_{\phi}/(3M_{\rm pl}^2)$. 
{}From Eq.~(\ref{beani2}) the slow-roll parameter 
$\epsilon=-\dot{H}/H^2$ is much smaller than 1
under the condition $\dot{\phi}^2/2 \ll V(\phi)$.
If the shear $\Sigma$ approaches a constant 
value, Eq.~(\ref{beani3}) shows that the ratio 
$\Sigma/H$ converges to 
\be
\frac{\Sigma}{H} \simeq \frac{2\rho_A}{3V}\,.
\label{sigH}
\ee
In order to keep the energy density $\rho_A$ 
nearly constant, we require that 
\be
f(\phi) \simeq e^{-2\alpha}=a^{-2}\,,
\label{fphi}
\ee
where we used the property $|\sigma| \ll \alpha$.
Neglecting the contribution of the vector field and the $\ddot{\phi}$
term on the l.h.s. of Eq.~(\ref{beani4}), we obtain $3 \dot{\alpha}\dot{\phi} \simeq -V_{,\phi}$ 
and hence $d\alpha/d\phi \simeq -V/(M_{\rm pl}^2V_{,\phi})$.
Integrating this equation, the critical coupling (\ref{fphi}) 
can be expressed as
\begin{equation}
f(\phi)=e^{2\int \frac{V}
{M_{\rm pl}^2 V_{,\phi}}d\phi}\,.
\label{fphive}
\end{equation}

Let us substitute Eq.~(\ref{fphive}) for Eq.~(\ref{beani4}) 
and drop the $\ddot{\phi}$ term alone, i.e., 
\begin{equation}
\frac{d\phi}{d\alpha} \simeq 
-\frac{M_{\rm pl}^2V_{,\phi}}{V}+
\frac{2p_A^2}{V_{,\phi}}
e^{-4\alpha-4\sigma
-4 \int \frac{V}{M_{\rm pl}^2 V_{,\phi}}d\phi}\,.
\end{equation}
Neglecting the variation of $\phi/M_{\rm pl}$ relative to 
that of $\alpha$, this equation can be integrated to give
\be
e^{4\alpha+4\sigma+4\int \frac{V}
{M_{\rm pl}^2 V_{,\phi}}d\phi} \simeq 
\frac{8p_A^2 V}{M_{\rm pl}^2 V_{,\phi}^2} 
(\alpha+\alpha_0)\,,
\label{intvec}
\ee
where $\alpha_0>0$ is an integration constant. 
Substituting Eqs.~(\ref{fphive}) and (\ref{intvec}) into 
Eq.~(\ref{rhoA}), we obtain the following relation 
\be
r_A \equiv \frac{\rho_A}{\epsilon V} \simeq \frac{1}{8(\alpha+\alpha_0)}\,,
\label{rA}
\end{equation}
where we have used the property
$\epsilon \simeq (M_{\rm pl}^2/2)(V_{,\phi}/V)^2$ 
under the slow-roll approximation.
As long as $\alpha \ll \alpha_0$, the ratio $r_A$ 
stays nearly constant. 
{}From Eq.~(\ref{sigH}) and (\ref{rA}) we have 
\be
\frac{\Sigma}{H} \simeq \frac{\epsilon}
{12(\alpha+\alpha_0)}\,,
\end{equation}
so that the anisotropic hair survives for $\alpha \ll \alpha_0$.

We can generalize the above discussion to the more general 
coupling $f(\phi)=e^{2c\int \frac{V}
{M_{\rm pl}^2 V_{,\phi}}d\phi}$, where $c$ is a constant.
When $c>1$, there is an attractor solution along which 
the anisotropic hair survives during inflation.
Along this attractor the shear to the Hubble parameter 
is given by $\Sigma/H \simeq (c-1)\epsilon/(3c)$ \cite{Watanabe}.

\subsection{Anisotropic power spectra}

For the anisotropic inflationary scenario described by 
the action (\ref{eq:action}), let us derive the scalar 
power spectrum in the form (\ref{anispe}).
Since the anisotropy of the expansion rate should be 
sufficiently small for compatibility with observations, we can 
ignore the effect of the anisotropic expansion for the derivation of 
the perturbation equations \cite{WataSoda}.  
Then we use the perturbed metric (\ref{permet}) with 
the curvature perturbation $\zeta$ defined in Eq.~(\ref{zeta}).
We choose the spatially flat gauge ($\psi=0$), so that 
$\zeta=-(H/\dot{\phi})\delta \phi$. 

The curvature perturbation is decomposed into the isotropic 
field $\zeta^{(0)}$ and the anisotropic contribution $\delta \zeta$, as 
$\zeta=\zeta^{(0)}+\delta \zeta$.
Decomposing $\zeta$ into the Fourier components as 
Eq.~(\ref{RFourier}), the solution to the isotropic component
$\zeta^{(0)}(\tau,k)$ is given by Eq.~(\ref{usol}) with 
$c_s^2=1$ and $Q_s=M_{\rm pl}^2 \epsilon$.
{}From Eq.~(\ref{scalarpower}) the isotropic scalar 
power spectrum is
\be
{\mathcal P}_{\zeta}^{(0)}=\frac{H^{2}}{8\pi^{2}
M_{\rm pl}^{2}\epsilon}\,,
\label{scalarpower2}
\ee
which is evaluated at $k=aH$.

In the Coulomb gauge, the vector field $A_{\mu}$ can be 
decomposed into the Fourier components,
\begin{eqnarray}
& & A_i({\bm x},\tau) = A_i^{(0)} (\tau)+ \delta A_i\,, \nonumber \\
&& \delta A_i=
\sum_{\lambda=1,2} \int \frac{d^3k}{(2\pi)^{3/2}} 
e^{i{\bm k} \cdot {\bm x}} \left[ 
A_{\lambda}(k, \tau)a_{\lambda}({\bm k}) 
+A_{\lambda}^* (k,\tau)a_{\lambda}^{\dagger} (-{\bm k}) \right]
{\epsilon}^{(\lambda)}_i ({\bm k})\,,
\end{eqnarray}
where $A_i^{(0)} (\tau)=(A_x^{(0)},0,0)$ is the background component, 
and ${\epsilon}^{(\lambda)}_i ({\bm k})$ ($\lambda=1,2$) 
are polarization vectors satisfying the relations
$k_i {\epsilon}^{(\lambda)}_i ({\bm k}) =0$, 
${\epsilon}^{(\lambda)}_i(-{\bm k}) ={ \epsilon}_i^{{*(\lambda)}}({\bm k})$, and
${ \epsilon}^{(\lambda)}_i ({\bm k}) \  { \epsilon}^{*{(\lambda ')}}_i ({\bm k})
=\delta_{\lambda\lambda '}$. Introducing a rescaled field
$V_{\lambda}=fA_{\lambda}$, it obeys the equation of motion 
\begin{equation}
V_{\lambda}''+\left(k^2 -\frac{f''}{f} \right)V_{\lambda}=0\,.
\label{Vlam}
\end{equation}

In the following, let us focus on the coupling (\ref{fphi}). 
Since $f \propto \tau^2$ on the de Sitter background ($a=-(\tau H)^{-1}$),
it follows that $f''/f=2/\tau^2$. In this case, the vector-field perturbation 
is scale-invariant. Imposing the Bunch-Davies vacuum 
in the asymptotic past, the solution to Eq.~(\ref{Vlam}) is given by 
\begin{equation}
A_{\lambda}(k, \tau)= 
\frac{Ha^3}{\sqrt{2k^3}}(1+ik \tau)e^{-i k\tau}\,.
\label{Alam}
\end{equation}

On the flat FLRW background with 
the line element $ds^2=a^2(-d\tau^2+\delta_{ij}dx^idx^j)$,
the tree-level interacting Lagrangian 
$L_{\rm int}=-\sqrt{-g}f^2(\phi)F_{\mu \nu}F^{\mu \nu}/4$ can be 
expanded up to second order in perturbations with the 
expansion $f^2=\bar{f}^2+(\partial \bar{f}^2/\partial \phi)\delta \phi$ 
and $F_{\mu \nu}=\bar{F}_{\mu \nu}+\delta F_{\mu \nu}$ (a bar 
represents background values). 
On using the property $(\partial \bar{f}^2/\partial \phi)\delta \phi=4f^2\zeta$ 
in the spatially flat gauge, the second-order interacting Lagrangian 
for curvature perturbations reads
\begin{equation}
L_{\rm int}^{(2)}=4a^4E_x \delta E_x \zeta\,,
\end{equation}
where $E_x=fA_{x}^{(0)'}/a^2$ and 
$\delta E_i=f \delta A_i'/a^2$. 
{}From (\ref{Alam}) the solution to $\delta E_i ({\bm x},\tau)$ 
in the super-Hubble regime ($|k \tau| \ll 1$) can be expressed as
\be
\delta E_i ({\bm x}, \tau)=\int \frac{d^3k}{(2\pi)^{3/2}} e^{i {\bm k} \cdot {\bm x}}
\delta \hat{E}_i ({\bm k}, \tau),\qquad
\delta \hat{E}_i ({\bm k}, \tau)=
\sum_{\lambda=1,2} \frac{3H^2}{\sqrt{2k^3}} 
\left[ a_{\lambda} ({\bm k})+a_{\lambda}^{\dagger} (-{\bm k}) 
\right] {\epsilon}^{(\lambda)}_i ({\bm k}).
\label{delEi}
\ee
Then, the interacting Hamiltonian $H_{\zeta}=- \int d^3x\,L_{\rm int}^{(2)}$ 
is given by 
\be
H_{\zeta}=-\frac{4E_x}{H^4 \tau^4} \int d^3k\, 
\delta \hat{E}_x({\bm k}, \tau) \hat{\zeta}^{(0)}(-{\bm k}, \tau)\,.
\label{Hint}
\ee
{}From this we can compute the contribution to 
the two-point correlation function of scalar perturbations, as 
\begin{eqnarray}
\delta \langle 0|\hat{\zeta}({\bm k}_1) \hat{\zeta}({\bm k}_2) |0 \rangle 
&=& 
- \int_{\tau_{{\rm min},1}}^\tau  d\tau_1 \int_{\tau_{{\rm min},2}}^{\tau_1} d\tau_2\, \langle 0| 
\left[ \left[ \hat{\zeta}^{(0)} ({\bm k}_1, \tau) 
\hat{\zeta}^{(0)} ({\bm k}_2, \tau) , 
H_\zeta (\tau_1 ) \right] , H_\zeta (\tau_2) \right] |0 \rangle  \nonumber\\
&=& \frac{4E_{x}^2}{9\epsilon^2 M_{\rm pl}^4 H^4} \prod_{i=1}^2 
\int_{-1/k_i}^{\tau} \frac{d\tau_i}{\tau_i^4} \left( \tau^3 - \tau_i^3 \right) 
\langle 0|\delta \hat{E}_{x}({\bm k}_1, \tau_1) \delta {\hat{E}}_{x}
({\bm k}_2, \tau_2) |0 \rangle  \nonumber\\
&=&  \frac{2\pi^2}{k_1^3} \delta^{(3)} ({\bm k}_1 + {\bm k}_2 )  
\frac{E_{x}^2N_k^2 \sin^2 \theta_{{\bm k}_1, {\bm x}}}{\pi^2 \epsilon^2 M_{\rm pl}^4}\,,
\label{powerspe}
\end{eqnarray}
where we have used the property 
\begin{equation}
[\hat{\zeta}^{(0)}({\bm k},\tau), \hat{\zeta}^{(0)}({\bm k'},\tau')]
\simeq -i \frac{H^2}{6\epsilon M_{\rm pl}^2} (\tau^3-\tau{'^3})
\delta^{(3)} ({\bm k}+{\bm k}')\,.
\end{equation}
In the first line of Eq.~(\ref{powerspe}) we have evaluated  
the two integrals in the super-horizon regime ($-k_i \tau<1$), 
i.e.,  $\tau_{{\rm min},i}=-1/k_i$ with $i=1,2$. 
We also employed the relation $\int_{-1/k_i}^{\tau}d\tau_i\,(\tau^3-\tau_i^3)/\tau_i^4
\simeq \ln (aH/k_i) \simeq N_{k_i}$ in the regime $-k_i \tau \ll 1$, where
$N_{k_i}$ is the number of e-foldings before the end of 
inflation at which the modes with the wavenumber $k_i$ 
left the Hubble radius. 
Since ${\bm k}_1=-{\bm k}_2$, it follows that 
$N_{k_1}=N_{k_2} \equiv N_k$.

{}From Eqs.~(\ref{scalarpower2}) and (\ref{powerspe}), the total 
scalar power spectrum is given by 
\be
{\mathcal P}_{\zeta}={\mathcal P}_{\zeta}^{(0)}
\left( 1+48 r_A N_k^2  \sin^2 
\theta_{{\bm k}_1,{\bm x}} \right) \,,
\label{Pfull}
\ee
where we have used the relation $\rho_A=E_x^2/2$ and 
the definition $r_A$ given in Eq.~(\ref{rA}).
Comparing the spectrum (\ref{Pfull}) with the parametrization 
(\ref{anispe}), it follows that 
\be
g_*=-48\,r_A N_k^2=-48\frac{\rho_A}{\epsilon V}N_k^2\,.
\ee
Since $g_*<0$, the power spectrum has an oblate-type anisotropy.
The condition $|g_*| \lesssim 0.01$ translates to 
$r_A \lesssim 10^{-7}$ for $N_k \sim 60$. 
In Eq.~(\ref{rA}) the parameter $\alpha$ corresponds to the number of e-foldings 
from the onset of inflation, so that $r_A \simeq 1/(8\alpha_0)=$\,constant 
for $\alpha \ll 10^6$. Thus, the model (\ref{eq:action}) with 
the coupling (\ref{fphi}) can explain the broken rotational 
invariance of the scalar power spectrum.

The tensor power spectrum can be computed in 
a similar way from the interacting Hamiltonian with the vector 
field $A_i$ and the tensor perturbation $h_{ij}$. 
It is given by \cite{WataSoda,Ohashi2}
\be
{\mathcal P}_{h}=16\epsilon {\mathcal P}_{\zeta}^{(0)} 
\left( 1+12\epsilon r_A N_k^2  \sin^2 
\theta_{{\bm k}_1,{\bm x}} \right) \,.
\label{Ph_vec}
\ee
Compared to the scalar spectrum (\ref{Pfull}), the anisotropic 
contribution is suppressed due to the additional factor $\epsilon$.
Then the presence of anisotropies leads to the suppressed 
tensor-to-scalar ratio $r={\mathcal P}_h/{\mathcal P}_{\zeta}$.
For increasing $N_k$, the scalar amplitude ${\mathcal P}_{\zeta}$
gets larger, which leads to the decrease of $n_s$. 
If $|g_*|$ is larger than the order of 0.1, observational 
constraints on inflaton potentials in the ($n_s, r$) plane
are subject to change \cite{Ohashi2}.

\subsection{Primordial non-Gaussianities}

We also compute the three-point correlation function of curvature perturbations 
for the coupling (\ref{fphi}). 
In addition to the second-order interacting Lagrangian $L_{\rm int}^{(2)}$,
there is a contribution to the bispectrum coming from the third-order 
interacting Lagrangian $L_{\rm int}^{(3)}\simeq 2a^4\delta E_i \delta E_j \zeta$.
The corresponding interacting Hamiltonian is given by 
\begin{equation}
H_{\zeta2}=-\frac{2}{H^4\tau^4}\int \, \frac{d^3k d^3p}{(2\pi)^{3/2}} \, 
\delta \hat{E}_i({\bm k},\tau) \delta \hat{E}_j({\bm p},\tau) 
\hat{\zeta}^{(0)}(-{\bm k}-{\bm p},\tau) \,.
\label{Hzeta3}
\end{equation}
Then the three-point correlation of $\zeta$ can be evaluated as \cite{Bartolo}
\begin{eqnarray}
\hspace{-1.2cm}& &
\delta \langle 0|\hat{\zeta}({\bm k}_1) \hat{\zeta}({\bm k}_2)\hat{\zeta}({\bm k}_3) |0 \rangle 
= i \int_{\tau_{{\rm min},1}}^\tau  d\tau_1 \int_{\tau_{{\rm min},2}}^{\tau_1} 
d\tau_2\, \int_{\tau_{{\rm min},3}}^{\tau_2} d\tau_3\, \nonumber \\
\hspace{-1.2cm}&& \times \langle 0| 
\left[\left[ \left[ \hat{\zeta}^{(0)} ({\bm k}_1) 
\hat{\zeta}^{(0)} ({\bm k}_2) \hat{\zeta}^{(0)} ({\bm k}_3)(\tau) , 
H_{\zeta2} (\tau_1 ) \right] , H_\zeta (\tau_2) \right] H_\zeta (\tau_3)  
\right] |0 \rangle 
+ \text{2~permutations}  \nonumber\\
\hspace{-1.2cm}&& \simeq 288\sqrt{2}\pi^{5/2} \frac{E_{x}^2}{\epsilon V} 
N_{k_1} N_{k_2} N_{k_3} 
({\mathcal P}_{\zeta}^{(0)})^2 \delta^{(3)} ({\bm k}_1 + {\bm k}_2 + {\bm k}_3)
 \nonumber \\
\hspace{-1.2cm}&&\times \left[\frac{1}{k_1^3 k_2^3}( 1- \cos^2 \theta_{{\bm k}_1, {\bm x}}
-\cos^2\theta_{{\bm k}_2, {\bm x}}+ \cos\theta_{{\bm k}_1,  {\bm x}}\cos\theta_{{\bm k}_2, 
{\bm x}}\cos\theta_{{\bm k}_1, {\bm k}_2})
+{\rm 2~permutations}\right].
\label{bispe}
\end{eqnarray}
The nonlinear parameter $f_{\rm NL}$ following from 
the bispectrum (\ref{bispe}) reads \cite{Bartolo}
\begin{eqnarray}
\hspace{-0.7cm}
f_{\rm NL} &=&
6 \left(\frac{-g_*}{0.01}\right)\left(\frac{N_k}{60}\right)\frac{1}{k_1^3+k_2^3+k_3^3} 
[k_3^3( 1- \cos^2 \theta_{{\bm k}_1, {\bm x}} -\cos^2\theta_{{\bm k}_2, {\bm x}} \nonumber \\
& &+ \cos\theta_{{\bm k}_1,  {\bm x}}\cos\theta_{{\bm k}_2,  {\bm x}}
\cos\theta_{{\bm k}_1, {\bm k}_2})
+{\rm 2~permutations}]\,,
\label{fnl_vec}
\end{eqnarray}
where we have used the approximations 
$({\mathcal P}_{\zeta})^2\simeq ({\mathcal P}_{\zeta}^{(0)})^2$ and 
$N_{k_1}\simeq N_{k_2}\simeq N_{k_3} \equiv N_k$.

In the strict squeezed limit characterized by $k_3 \to 0$ and 
$\theta_{{\bm k}_1, {\bm k}_2} \to \pi$, the nonlinear estimator $f_{\rm NL}$ 
vanishes for any values of $\theta_{{\bm k}_1,  {\bm x}}$ \cite{Ohashi2}.
This corresponds to the case in which the angles 
$\theta_{{\bm k}_2, {\bm k}_3}$ and $\theta_{{\bm k}_3, {\bm k}_1}$ approach $\pi/2$.
For the incomplete squeezed shape where the angles $\theta_{{\bm k}_2 ,{\bm k}_3}$ and 
$\theta_{{\bm k}_3 ,{\bm k}_1}$ are not necessarily close to $\pi /2$, 
we can take any angle between ${\bm k}_3$ and ${\bm k}_1, {\bm k}_2$. 
Averaging over $f_{\rm NL}$ in all the directions, the nonlinear estimator 
of the squeezed shape ($k_3\ll k_1\simeq k_2$, $\theta_{{\bm k}_1, {\bm k}_3}\to \pi
-\theta_{{\bm k}_2, {\bm k}_3}$, and $\theta_{{\bm k}_2, {\bm x}} \to \pi 
- \theta_{{\bm k}_1, {\bm x}}$) 
can be estimated as \cite{Bartolo}
\be
f_{\rm NL}^{\rm local, average} \simeq 2.7\left(\frac{-g_*}{0.01}\right)\left(\frac{N_k}{60}\right)
\frac{[1- \cos^2 \theta_{{\bm k}_1, {\bm x}} -\cos^2\theta_{{\bm k}_3,{\bm x}}
+ \cos\theta_{{\bm k}_1, {\bm x}}\cos\theta_{{\bm k}_3, {\bm x}}\cos\theta_{{\bm k}_1, {\bm k}_3}]}{4/9} \,,
\label{fnl_vec_local}
\ee
where we have used the fact that the average of the function 
in the last square bracket integrated over all the angles is $4/9$.
If $g_*=-0.01$ and $N=60$, then $f_{\rm NL}^{\rm local, average}=2.7$ and hence 
the model can be compatible with the Planck bound 
$f_{\rm NL}^{\rm local}=2.7 \pm 5.8$ (68\,\%\,CL).
In the equilateral limit ($k_1=k_2=k_3$) the non-linear 
estimator (\ref{fnl_vec}) reduces to
\be
f_{\rm NL}^{\rm equil} \simeq 0.75 \left( \frac{-g_*}{0.01} \right)
\left( \frac{N_k}{60} \right)\,, 
\ee
which is smaller than the order of 1 for $|g_*|<0.01$.

\subsection{Generality of anisotropic inflation}

So far we have focused on the case of potential-driven anisotropic 
slow-roll inflation, but the anisotropic hair can also survive in other 
inflationary scenarios. 
For example, in k-inflation, the power-law cosmic acceleration 
($a \propto t^p$ with $p>1$) can be realized for the Lagrangian 
$P=Xg(Y)$ \cite{ghost2,tsuasi}, where $g$ is an arbitrary function 
in terms of $Y=Xe^{\lambda \phi/M_{\rm pl}}$ ($\lambda$ is a constant).
If the inflaton couples to the vector field $A_{\mu}$ with an exponential 
coupling $f(\phi) \propto e^{\mu \phi/M_{\rm pl}}$, the models 
with the Lagrangian $P=Xg(Y)$ give rise to anisotropic 
inflationary solutions with $\Sigma/H={\rm constant}$ \cite{anikinf}.
Moreover, it has also been shown that these anisotropic solutions are stable 
attractors irrespective of the forms of $g(Y)$, provided that they exist in the 
regime $\Sigma/H \ll 1$. This shows the generality of anisotropic inflation.
 
If the inflaton couples to a two-form field $B_{\mu \nu}$, the anisotropic 
hair can also survive during inflation \cite{Ohashi}. 
In such models the sign of $g_*$ is positive, so the scalar power 
spectrum has a prolate-type anisotropy. 
The effect of anisotropies appears in a similar way 
to the scalar and tensor power spectra, i.e., 
both $n_s$ and $r$ get smaller for larger $g_*$ \cite{Ohashi2}.
The non-linear estimator in the two-form field model is generally smaller 
than that in the vector model for the same orders of $|g_*|$, so 
the former is even more likely to satisfy the Planck bounds of 
non-Gaussianities. In the two-form field model there is no cross correlation 
between scalar and tensor perturbations \cite{Ohashi2}, while 
in the vector model the cross correlation does not vanish \cite{WataSoda}.
Hence these two models can be distinguished observationally.
We refer the reader to Refs.~\cite{Ohashi,Ohashi2}
for detailed calculations of the inflationary observables.

\section{Conclusion}
\label{consec}

In this review we have constrained a host of inflationary 
models by using the Planck data combined with other observations. 
In particular, most single-field inflationary models proposed in 
the literature belong to a class of the Horndeski theory described 
by the action (\ref{action}). 
We have computed the power spectra of scalar and tensor 
perturbations in such general theories to confront each model 
with observations of CMB temperature anisotropies. 
Since the non-linear estimator $f_{\rm NL}$ of scalar non-Gaussianities 
in the squeezed limit is much smaller than 1 under the slow-variation 
approximation, the models based on the Horndeski theory are compatible 
with the recent Planck bound.

We have applied our general results to concrete models of inflation 
such as potential-driven slow-roll inflation, k-inflation, Starobinsky inflation, 
and Higgs inflation with non-minimal/derivative/Galileon couplings. 
In the potential-driven slow-roll scenario, models such as hybrid inflation 
($V(\phi)=\Lambda^4+m^2 \phi^2/2$) and power-law inflation 
($V(\phi)=V_0e^{-\gamma \phi/M_{\rm pl}}$) are outside the 
95\,\%\,CL boundary constrained by 
Planck+WP+BAO+high-$\ell$.
The monomial potential $V(\phi)=\lambda_n \phi^n/n$ ($n>0$) 
is outside the 68\,\%\,CL region.
In natural inflation with the potential $V(\phi)=V_0[1+\cos(\phi/f)]$
the symmetry-breaking scale $f$ is constrained as 
$5.1M_{\rm pl}<f<7.9M_{\rm pl}$ (68\,\%\,CL).
Very small-field potentials such as $V(\phi)=\Lambda^4 (1-e^{-\phi/M})$
are consistent with the data because of the suppressed tensor-to-scalar ratio.

K-inflation can be tightly constrained by adding the bound on 
the equilateral non-linear estimator $f_{\rm NL}^{\rm equil}$
to those of $n_s$ and $r$. In the dilatonic ghost condensate model described
by the Lagrangian (\ref{dila}), the scalar propagation speed is constrained
as $0.034<c_s<0.046$ (95\,\%\,CL) from the bounds of $n_s$ and $r$, 
but, in this parameter range, $|f_{\rm NL}^{\rm equil}|$ is too large to 
be compatible with the Planck data. 
The same property also holds for the ultraviolet DBI model.
In the infrared DBI model the allowed parameter space satisfying 
all the bounds is constrained to a narrow range.

In Starobinsky inflation the scalar spectral index and the tensor-to-scalar 
ratio are given by $n_s=1-2/N$ and $r=12/N^2$ respectively, 
in which case the model is well within the 68\,\%\,CL region.
In Higgs inflation, described by the potential 
$V(\phi)=(\lambda_4/4)(\phi^2-v^2)^2$ ($v \sim 10^2$\,GeV),
the presence of non-minimal couplings $-\xi \phi^2 R/2$ 
with $|\xi| \gg 1$ gives rise to the Einstein-frame 
potential similar to that in Starobinsky inflation, so that 
$n_s$ and $r$ are the same in both models as long as 
quantum corrections to the tree-level Higgs potential are suppressed.
It is possible to realize the self coupling $\lambda_4$ of the order 
of 0.1 at the expense of having a large negative non-minimal 
coupling $\xi \sim -10^4$. 

In the presence of the field-derivative coupling to the Einstein tensor 
or the Galileon self-interactions, 
the Higgs potential is still outside the 68\,\%\,CL region. 
Although such couplings lead to the decrease of $r$ due to the enhanced 
friction for the inflaton, these models are not necessarily favored
over non-minimally coupled Higgs inflation or Starobinsky inflation
because of the tight upper bound on $n_s$ provided by the Planck data.

We have also reviewed anisotropic inflation driven by the presence 
of a coupling between a vector field and the inflaton.
For the coupling (\ref{fphi}) an anisotropic hair survives during inflation, 
so that several observational signatures can be imprinted in CMB. 
We have derived the anisotropy parameter $g_*$, appearing in the 
scalar power spectrum, and have also computed the bispectrum of 
primordial non-Gaussianities. Under the bound $|g_*|<0.01$
the non-linear parameter $f_{\rm NL}$ is smaller than the 
order of 1, in which case the Planck bound on
non-Gaussianities is satisfied. 
We also note that the anisotropic hair can survive for power-law 
k-inflation or in the presence of a coupling between 
a two-form field and the inflaton.

It is expected that future observations of CMB polarization 
such as LiteBIRD will provide further tight constraints on the 
amplitude of gravitational waves. 
We hope that we can approach the best model of inflation 
in the foreseeable future.

\section*{Acknowledgement}

The author is supported by the Scientific Research Fund of the 
JSPS (No.~24540286) and Scientific Research 
on Innovative Areas (No.~21111006).  



\begin{thebibliography}{9}

\bibitem{Sta80}
A.~A.~Starobinsky,
Phys.\ Lett.\ B {\bf 91}, 99 (1980).

\bibitem{oldinf}
D.~Kazanas,
Astrophys.\ J.\  {\bf 241} L59 (1980);
K.~Sato, Mon.\ Not.\ R.\ Astron.\ Soc. {\bf 195}, 467 (1981);
Phys.\ Lett.\ {\bf 99B}, 66 (1981);
A.~H.~Guth,
Phys.\ Rev.\ D {\bf 23}, 347 (1981).

\bibitem{oldper}
V.~F.~Mukhanov and G.~V.~Chibisov,
JETP Lett.\  {\bf 33}, 532 (1981);
A.~H.~Guth and S.~Y.~Pi,
Phys.\ Rev.\ Lett.\  {\bf 49} (1982) 1110;
S.~W.~Hawking,
Phys.\ Lett.\ B {\bf 115}, 295 (1982);
A.~A.~Starobinsky,
Phys.\ Lett.\ B {\bf 117} (1982) 175;
J.~M.~Bardeen, P.~J.~Steinhardt and M.~S.~Turner,
Phys.\ Rev.\ D {\bf 28}, 679 (1983).

\bibitem{COBE}
G.~F.~Smoot {\it et al.},
Astrophys.\ J.\  {\bf 396}, L1 (1992).

\bibitem{WMAP1}
D.~N.~Spergel {\it et al.}  [WMAP Collaboration],
Astrophys.\ J.\ Suppl.\  {\bf 148}, 175 (2003).

\bibitem{Planck}
P.~A.~R.~Ade {\it et al.}  [Planck Collaboration],
arXiv:1303.5076 [astro-ph.CO].

\bibitem{Starreheating}
A.~A.~Starobinsky, 
``Nonsingular model of the Universe with the quantum-gravitational de Sitter stage and its observational consequences'', in: Proc. of the 2nd Seminar, ``Quantum Gravity''
(Moscow, 13-15 Oct. 1981), INR Press, Moscow, 1982, pp. 58-72, M.~A.~Markov, P.~C.~West (eds.) Publ. Co., New York, 1984, pp. 103-128); 
A.~Vilenkin, 
Phys.\ Rev.\  D {\bf 32}, 2511 (1985);
M.~B.~Mijic, M.~S.~Morris and W.~M.~Suen,
Phys.\ Rev.\  D {\bf 34}, 2934 (1986).

\bibitem{Linde82}
A.~D.~Linde,
Phys.\ Lett.\ B {\bf 108}, 389 (1982).

\bibitem{Albrecht}
A.~Albrecht and P.~Steinhardt,
Phys.~Rev.~Lett. {\bf 48}, 1220 (1982).

\bibitem{LRreview}
D.~H.~Lyth and A.~Riotto,
Phys.\ Rept.\  {\bf 314}, 1 (1999).

\bibitem{Lindebook}
A.~D.~Linde,
``Particle Physics and Inflationary Cosmology,''
arXiv:hep-th/0503203.

\bibitem{Bau}
D.~Baumann and L.~McAllister,
Ann.\ Rev.\ Nucl.\ Part.\ Sci.\  {\bf 59}, 67 (2009).

\bibitem{Mazumdar}
A.~Mazumdar and J.~Rocher,
Phys.\ Rept.\  {\bf 497}, 85 (2011).

\bibitem{Linde83}
A.~D.~Linde,
Phys.\ Lett.\ B {\bf 129} 177 (1983).

\bibitem{Kolb}
J.~E.~Lidsey, A.~R.~Liddle, E.~W.~Kolb, E.~J.~Copeland,
Rev.\ Mod.\ Phys.\  {\bf 69}, 373 (1997).

\bibitem{LLbook}
A.~R.~Liddle and D.~H. ~Lyth, 
{\em Cosmological inflation and large-scale structure}, 
Cambridge University Press (2000).

\bibitem{BTW}
B.~A.~Bassett, S.~Tsujikawa and D.~Wands,
Rev.\ Mod.\ Phys.\  {\bf 78}, 537 (2006).

\bibitem{Martin} 
J.~Martin, C.~Ringeval and V.~Vennin,
arXiv:1303.3787 [astro-ph.CO].

\bibitem{Adeinf} 
P.~A.~R.~Ade {\it et al.}  [Planck Collaboration],
arXiv:1303.5082 [astro-ph.CO].

\bibitem{Kuro} 
S.~Tsujikawa, J.~Ohashi, S.~Kuroyanagi and A.~De Felice,
Phys.\ Rev.\ D {\bf 88}, 023529 (2013).

\bibitem{WMAP9} 
G.~Hinshaw {\it et al.}  [WMAP Collaboration],
Astrophys.\ J.\ Suppl.\  {\bf 208}, 19 (2013).

\bibitem{BAO} 
F.~Beutler {\it et al.},
Mon.\ Not.\ Roy.\ Astron.\ Soc.\  {\bf 416}, 3017 (2011); 
N.~Padmanabhan  {\it et al.},
arXiv:1202.0090 [astro-ph.CO];
L.~Anderson {\it et al.},
Mon.\ Not.\ Roy.\ Astron.\ Soc.\  {\bf 427}, no. 4, 3435 (2013).

\bibitem{highl} 
C.~L.~Reichardt  {\it et al.},
Astrophys.\ J.\  {\bf 755}, 70 (2012);
S.~Das {\it et al.},
arXiv:1301.1037 [astro-ph.CO].

\bibitem{kinflation}
C.~Armendariz-Picon, T.~Damour and V.~F.~Mukhanov,
Phys.\ Lett.\  B {\bf 458}, 209 (1999).

\bibitem{Garriga}
J.~Garriga and V.~F.~Mukhanov,
Phys.\ Lett.\  B {\bf 458}, 219 (1999).

\bibitem{Seery}
D.~Seery and J.~E.~Lidsey,
JCAP {\bf 0506}, 003 (2005).

\bibitem{Chen}
X.~Chen, M.~x.~Huang, S.~Kachru and G.~Shiu,
JCAP {\bf 0701}, 002 (2007).

\bibitem{Adenon}
P.~A.~R.~Ade {\it et al.}  [Planck Collaboration],
arXiv:1303.5084 [astro-ph.CO].

\bibitem{FM} 
T.~Futamase and K.~-i.~Maeda,
Phys.\ Rev.\ D {\bf 39}, 399 (1989);
R.~Fakir and W.~G.~Unruh,
Phys.\ Rev.\ D {\bf 41}, 1783 (1990).

\bibitem{Higgs} 
F.~L.~Bezrukov and M.~Shaposhnikov,
Phys.\ Lett.\ B {\bf 659}, 703 (2008).

\bibitem{Brans} 
C.~Brans and R.~H.~Dicke,
Phys.\ Rev.\  {\bf 124}, 925 (1961).

\bibitem{Nico} 
A.~Nicolis, R.~Rattazzi and E.~Trincherini,
Phys.\ Rev.\ D {\bf 79}, 064036 (2009).

\bibitem{Vik} 
C.~Deffayet, G.~Esposito-Farese and A.~Vikman,
Phys.\ Rev.\ D {\bf 79}, 084003 (2009);
C.~Deffayet, S.~Deser and G.~Esposito-Farese,
Phys.\ Rev.\ D {\bf 80}, 064015 (2009).

\bibitem{KobaGa} 
T.~Kobayashi, M.~Yamaguchi and J.~'i.~Yokoyama,
Phys.\ Rev.\ Lett.\  {\bf 105}, 231302 (2010).

\bibitem{Galileons} 
C.~Burrage, C.~de Rham, D.~Seery and A.~J.~Tolley,
JCAP {\bf 1101}, 014 (2011).

\bibitem{Amendola} 
L.~Amendola,
Phys.\ Lett.\ B {\bf 301}, 175 (1993).

\bibitem{Germani} 
C.~Germani and A.~Kehagias,
Phys.\ Rev.\ Lett.\  {\bf 105}, 011302 (2010);
C.~Germani and A.~Kehagias,
Phys.\ Rev.\ Lett.\  {\bf 106}, 161302 (2011).

\bibitem{Nakayama} 
K.~Nakayama and F.~Takahashi,
JCAP {\bf 1011}, 009 (2010)
[arXiv:1008.2956 [hep-ph]].

\bibitem{Reza} 
A.~De Felice, S.~Tsujikawa, J.~Elliston and R.~Tavakol,
JCAP {\bf 1108}, 021 (2011).

\bibitem{Horndeski} 
G.~W.~Horndeski, Int.\ J.\ Theor.\ Phys.\ 10, 363-384 (1974).

\bibitem{DGSZ} 
C.~Deffayet, X.~Gao, D.~A.~Steer and G.~Zahariade,
Phys.\ Rev.\ D \textbf{84}, 064039 (2011);
C.~Charmousis, E.~J.~Copeland, A.~Padilla and P.~M.~Saffin, 
Phys.\ Rev.\ Lett.\ \textbf{108}, 051101 (2012).

\bibitem{Koba11} 
T.~Kobayashi, M.~Yamaguchi and J.~'i.~Yokoyama,
Prog.\ Theor.\ Phys.\  {\bf 126}, 511 (2011).

\bibitem{Gao} 
X.~Gao and D.~A.~Steer,
JCAP {\bf 1112}, 019 (2011).

\bibitem{DT} 
A.~De Felice and S.~Tsujikawa,
Phys.\ Rev.\ D {\bf 84}, 083504 (2011).

\bibitem{Kobaten} 
X.~Gao, T.~Kobayashi, M.~Yamaguchi and J.~'i.~Yokoyama,
Phys.\ Rev.\ Lett.\  {\bf 107}, 211301 (2011).

\bibitem{Gao2} 
X.~Gao {\it et al.,}
PTEP {\bf 2013}, 053E03 (2013).

\bibitem{nonGa}
S.~Mizuno and K.~Koyama,
Phys.\ Rev.\ D {\bf 82}, 103518 (2010);
A.~De Felice and S.~Tsujikawa,
JCAP {\bf 1104}, 029 (2011);
T.~Kobayashi, M.~Yamaguchi and J.~'i.~Yokoyama,
Phys.\ Rev.\ D {\bf 83}, 103524 (2011).

\bibitem{aniobser1} 
N.~E.~Groeneboom and H.~K.~Eriksen,
Astrophys.\ J.\  {\bf 690}, 1807 (2009);
D.~Hanson and A.~Lewis,
Phys.\ Rev.\ D {\bf 80}, 063004 (2009);
L.~Ackerman, S.~M.~Carroll and M.~B.~Wise,
Phys.\ Rev.\ D {\bf 75}, 083502 (2007).

\bibitem{aniobser2} 
N.~E.~Groeneboom, L.~Ackerman, I.~K.~Wehus and H.~K.~Eriksen,
Astrophys.\ J.\  {\bf 722}, 452 (2010).

\bibitem{Rub} 
S.~R.~Ramazanov and G.~Rubtsov,
arXiv:1311.3272 [astro-ph.CO].

\bibitem{Kim} 
J.~Kim and E.~Komatsu,
Phys.\ Rev.\ D {\bf 88}, 101301 (2013).

\bibitem{Watanabe}
M.~a.~Watanabe, S.~Kanno and J.~Soda,
Phys.\ Rev.\ Lett.\  {\bf 102}, 191302 (2009).

\bibitem{Sodareview}
J.~Soda,
Class.\ Quant.\ Grav.\  {\bf 29}, 083001 (2012);
A.~Maleknejad, M.~M.~Sheikh-Jabbari and J.~Soda,
Phys.\ Rept.\  {\bf 528}, 161 (2013).

\bibitem{Gum} 
A.~E.~Gumrukcuoglu, B.~Himmetoglu and M.~Peloso,
Phys.\ Rev.\ D {\bf 81}, 063528 (2010);
T.~R.~Dulaney and M.~I.~Gresham,
Phys.\ Rev.\ D {\bf 81}, 103532 (2010).

\bibitem{WataSoda} 
M.~a.~Watanabe, S.~Kanno and J.~Soda,
Prog.\ Theor.\ Phys.\  {\bf 123}, 1041 (2010).

\bibitem{Ohashi} 
J.~Ohashi, J.~Soda and S.~Tsujikawa,
Phys.\ Rev.\ D {\bf 87}, 083520 (2013).

\bibitem{Bartolo} 
N.~Bartolo, S.~Matarrese, M.~Peloso and A.~Ricciardone,
Phys.\ Rev.\ D {\bf 87}, 023504 (2013).

\bibitem{Shiraishi} 
M.~Shiraishi, E.~Komatsu, M.~Peloso and N.~Barnaby,
JCAP {\bf 1305}, 002 (2013).

\bibitem{Ohashi2} 
J.~Ohashi, J.~Soda and S.~Tsujikawa,
JCAP {\bf 1312}, 009 (2013).

\bibitem{Leach} 
A.~R.~Liddle and S.~M.~Leach,
Phys.\ Rev.\ D {\bf 68}, 103503 (2003).

\bibitem{ADM} 
R.~L.~Arnowitt, S.~Deser and C.~W.~Misner, 
Phys. \ Rev. \ {\bf117}, 1595 (1960).

\bibitem{Maldacena} 
J.~M.~Maldacena, 
JHEP \textbf{0305}, 013 (2003).

\bibitem{Koyama} 
K.~Koyama,
Class.\ Quant.\ Grav.\  {\bf 27}, 124001 (2010).

\bibitem{Lukash}
V.~N.~Lukash,
Sov.\ Phys.\ JETP {\bf 52}, 807 (1980).

\bibitem{Bardeen}
J.~M.~Bardeen,
Phys.\ Rev.\ D {\bf 22}, 1882 (1980).

\bibitem{Kodama}
H.~Kodama and M.~Sasaki,
Prog.\ Theor.\ Phys.\ Suppl.\  {\bf 78}, 1 (1984);
V.~F.~Mukhanov, H.~Feldman and R.~Brandenberger,
Phys.\ Rept.\  {\bf 215}, 203 (1992);
K.~A.~Malik and D.~Wands,
Phys.\ Rept.\  {\bf 475}, 1 (2009).

\bibitem{KSpergel} 
E.~Komatsu and D.~N.~Spergel, 
Phys.\ Rev.\ \textbf{D63}, 063002 (2001).

\bibitem{Bartolo2} 
N.~Bartolo, S.~Matarrese, A.~Riotto, 
Phys.\ Rev.\ \textbf{D65}, 103505 (2002); 
N.~Bartolo, E.~Komatsu, S.~Matarrese and A.~Riotto, 
Phys.\ Rept.\ \textbf{402}, 103 (2004).

\bibitem{DT13} 
A.~De Felice and S.~Tsujikawa,
JCAP {\bf 1303}, 030 (2013).

\bibitem{Cremi} 
P.~Creminelli and M.~Zaldarriaga, 
JCAP \textbf{0410}, 006 (2004).

\bibitem{violation} 
F.~Arroja, A.~E.~Romano and M.~Sasaki,
Phys.\ Rev.\ D {\bf 84}, 123503 (2011); 
P.~Adshead, C.~Dvorkin, W.~Hu and E.~A.~Lim,
Phys.\ Rev.\ D {\bf 85}, 023531 (2012);
X.~Chen, H.~Firouzjahi, M.~H.~Namjoo and M.~Sasaki,
Europhys.\ Lett.\  {\bf 102}, 59001 (2013).

\bibitem{Mangano} 
G.~Mangano {\it et al}.,
Nucl.\ Phys.\ B {\bf 729}, 221 (2005).

\bibitem{Ichi} 
K.~Ichikawa and T.~Takahashi,
Phys.\ Rev.\ D {\bf 73}, 063528 (2006).

\bibitem{Maartens}
L.~Randall and R.~Sundrum,
Phys.\ Rev.\ Lett.\  {\bf 83}, 4690 (1999);
R.~Maartens, D.~Wands, B.~A.~Bassett and I.~Heard,
Phys.\ Rev.\ D {\bf 62}, 041301 (2000).

\bibitem{Ho}
R.~Brandenberger and P.~-M.~Ho,
Phys.\ Rev.\ D {\bf 66}, 023517 (2002).

\bibitem{Calcagni}
G.~Calcagni, S.~Kuroyanagi, J.~Ohashi and S.~Tsujikawa,
JCAP {\bf 1403}, 052 (2014).
  
\bibitem{Mca} 
L.~McAllister, E.~Silverstein and A.~Westphal,
Phys.\ Rev.\ D {\bf 82}, 046003 (2010).

\bibitem{West} 
E.~Silverstein and A.~Westphal,
Phys.\ Rev.\ D {\bf 78}, 106003 (2008).

\bibitem{natural}
K.~Freese, J.~A.~Frieman and A.~V.~Olinto,
Phys.\ Rev.\ Lett.\  {\bf 65}, 3233 (1990);\\
F.~C.~Adams, J.~R.~Bond, K.~Freese, J.~A.~Frieman and A.~V.~Olinto,
Phys.\ Rev.\ D {\bf 47}, 426 (1993).

\bibitem{hybrid} 
A.~D.~Linde,
Phys.\ Rev.\ D {\bf 49}, 748 (1994).

\bibitem{Dvali} 
G.~R.~Dvali, Q.~Shafi and R.~K.~Schaefer,
Phys.\ Rev.\ Lett.\  {\bf 73}, 1886 (1994).

\bibitem{Lyth} 
D.~H.~Lyth,
Phys.\ Rev.\ Lett.\  {\bf 78}, 1861 (1997).

\bibitem{Lyth2} 
D.~H.~Lyth,
Lect.\ Notes Phys.\  {\bf 738}, 81 (2008)
[hep-th/0702128].

\bibitem{Tye} 
G.~R.~Dvali and S.~H.~H.~Tye,
Phys.\ Lett.\ B {\bf 450}, 72 (1999).

\bibitem{Conlon}
J.~P.~Conlon and F.~Quevedo,
JHEP {\bf 0601}, 146 (2006).

\bibitem{KKLT}
S.~Kachru {\it et al.},
JCAP {\bf 0310}, 013 (2003);
J.~J.~Blanco-Pillado {\it et al.},
JHEP {\bf 0609}, 002 (2006);
D.~Baumann, A.~Dymarsky, I.~R.~Klebanov and L.~McAllister,
JCAP {\bf 0801}, 024 (2008);
S.~Panda, M.~Sami and S.~Tsujikawa,
Phys.\ Rev.\ D {\bf 76}, 103512 (2007).

\bibitem{ghost2}
F.~Piazza and S.~Tsujikawa,
JCAP {\bf 0407}, 004 (2004).

\bibitem{ghost}
N.~Arkani-Hamed, H.~C.~Cheng, M.~A.~Luty and S.~Mukohyama,
JHEP {\bf 0405}, 074 (2004);
N.~Arkani-Hamed, P.~Creminelli, S.~Mukohyama and M.~Zaldarriaga,
JCAP {\bf 0404}, 001 (2004).

\bibitem{DBIinf}
E.~Silverstein and D.~Tong,
Phys.\ Rev.\  D {\bf 70}, 103505 (2004).

\bibitem{DBIsky}
M.~Alishahiha, E.~Silverstein and D.~Tong,
Phys.\ Rev.\  D {\bf 70}, 123505 (2004).

\bibitem{Baumann} 
D.~Baumann and L.~McAllister,
Phys.\ Rev.\ D {\bf 75}, 123508 (2007);
J.~E.~Lidsey and I.~Huston,
JCAP {\bf 0707}, 002 (2007).

\bibitem{IR} 
X.~Chen,
Phys.\ Rev.\ D {\bf 71}, 063506 (2005).

\bibitem{assistedk} 
J.~Ohashi and S.~Tsujikawa,
Phys.\ Rev.\ D {\bf 83}, 103522 (2011).

\bibitem{Ohanlon}
J.~O'Hanlon, Phys.\ Rev.\ Lett.\ {\bf 29}, 137 (1972); 
T.~Chiba,
Phys.\ Lett.\  B {\bf 575}, 1 (2003).

\bibitem{Maeda} 
K.~-i.~Maeda,
Phys.\ Rev.\ D {\bf 39}, 3159 (1989).

\bibitem{equiva} 
N.~Makino and M.~Sasaki, 
Prog.\ Theor.\ Phys.\ \textbf{86}, 103 (1991); 
D.~I.~Kaiser, 
Phys.\ Rev.\ D \textbf{52}, 4295 (1995); 
A.~De Felice and S.~Tsujikawa,
Living Rev.\ Rel.\  {\bf 13}, 3 (2010).

\bibitem{fRper}
L.~A.~Kofman, V.~F.~Mukhanov and D.~Y.~.Pogosian,
Sov.\ Phys.\ JETP {\bf 66}, 433 (1987);
J.~-c.~Hwang and H.~Noh,
Phys.\ Lett.\ B {\bf 506}, 13 (2001).

\bibitem{Stainf}
S.~V.~Ketov and A.~A.~Starobinsky,
Phys.\ Rev.\ D {\bf 83}, 063512 (2011);
S.~V.~Ketov and S.~Tsujikawa,
Phys.\ Rev.\ D {\bf 86}, 023529 (2012);
J.~Ellis, D.~V.~Nanopoulos and K.~A.~Olive,
Phys.\ Rev.\ Lett.\  {\bf 111}, 111301 (2013);
Y.~Watanabe and J.~'i.~Yokoyama,
Phys.\ Rev.\ D {\bf 87}, 103524 (2013);
R.~Kallosh and A.~Linde,
JCAP {\bf 1306}, 028 (2013);
R.~Kallosh and A.~Linde,
JCAP {\bf 1307}, 002 (2013);
F.~Farakos, A.~Kehagias and A.~Riotto,
Nucl.\ Phys.\ B {\bf 876}, 187 (2013);
W.~Buchmuller, V.~Domcke and K.~Kamada,
Phys.\ Lett.\ B {\bf 726}, 467 (2013);
F.~Briscese, A.~Marciano, L.~Modesto and E.~N.~Saridakis,
Phys.\ Rev.\ D {\bf 87}, 083507 (2013);
F.~Briscese, L.~Modesto and S.~Tsujikawa,
Phys.\ Rev.\ D {\bf 89}, 024029 (2014).

\bibitem{Gumjud} 
E.~Komatsu and T.~Futamase, 
Phys.\ Rev.\ D \textbf{59}, 064029 (1999);
S.~Tsujikawa and B.~Gumjudpai,
Phys.\ Rev.\ D {\bf 69}, 123523 (2004).

\bibitem{Germani2} 
C.~Germani, L.~Martucci and P.~Moyassari,
Phys.\ Rev.\ D {\bf 85}, 103501 (2012).

\bibitem{Berera} 
A.~Berera,
Phys.\ Rev.\ Lett.\  {\bf 75}, 3218 (1995);
S.~Bartrum {\it et al.},
arXiv:1307.5868 [hep-ph].

\bibitem{Wata} 
C.~Germani and Y.~Watanabe,
JCAP {\bf 1107}, 031 (2011).

\bibitem{tsujideri} 
S.~Tsujikawa,
Phys.\ Rev.\ D {\bf 85}, 083518 (2012).

\bibitem{Kamada} 
K.~Kamada, T.~Kobayashi, M.~Yamaguchi and J.~'i.~Yokoyama,
Phys.\ Rev.\ D {\bf 83}, 083515 (2011).

\bibitem{OhashiGa} 
J.~Ohashi and S.~Tsujikawa,
JCAP {\bf 1210}, 035 (2012).

\bibitem{tsuasi}
S.~Tsujikawa,
Phys.\ Rev.\ D {\bf 73}, 103504 (2006);
L.~Amendola, M.~Quartin, S.~Tsujikawa and I.~Waga,
Phys.\ Rev.\ D {\bf 74}, 023525 (2006).

\bibitem{anikinf}
J.~Ohashi, J.~Soda and S.~Tsujikawa,
Phys.\ Rev.\ D {\bf 88}, 103517 (2013).

\end{thebibliography}
\end{document}